\documentclass[floatfix,aps,twocolumn,superscriptaddress,prb]{revtex4-2}

\usepackage{graphicx}
\usepackage{dcolumn} 
\usepackage{bm}
\usepackage[T1]{fontenc}
\usepackage[latin9]{inputenc}
\usepackage{textcomp}
\usepackage{amsmath}
\usepackage{mathrsfs}
\usepackage{graphicx}
\usepackage{amssymb}
\usepackage{enumitem}

\usepackage{units}
\usepackage{xcolor}
\usepackage{float}
\usepackage[colorlinks,citecolor=blue,urlcolor=blue]{hyperref}
\usepackage{physics}

\newcommand{\e}{\text{e}}

\newcommand{\gsp}{\ensuremath{g^{(1)}}}
\newcommand{\gp}{\ensuremath{g^{(1)}_{\rm pair}}}

\begin{document}

\title{Competition between pair and single-particle superfluidity in bosonic quasi-flat bands: A Gaussian state approach}

\newcommand{\cenoli}[0]{Center for Nonlinear Phenomena and Complex Systems, Universit\'e Libre de Bruxelles, Campus Plaine, B-1050 Brussels, Belgium}
\newcommand{\solvay}[0]{International Solvay Institutes, 1050 Brussels, Belgium}
\newcommand{\lkb}[0]{Laboratoire Kastler Brossel, Coll\`ege de France, CNRS, ENS-Universit\'e PSL, Sorbonne Universit\'e, 11 Place Marcelin Berthelot, 75005 Paris, France}

\author{Maxime Burgher}
\affiliation{\cenoli}
\affiliation{\solvay}

\author{Simon Loddo}
\affiliation{\cenoli}

\author{Laurens Vanderstraeten}
\affiliation{\cenoli}

\author{Nathan Goldman}
\affiliation{\cenoli}
\affiliation{\solvay}
\affiliation{\lkb}

\author{Ivan Amelio}
\email{ivan.amelio@ulb.be}
\affiliation{\cenoli}
\affiliation{\solvay}

\begin{abstract} 
The interplay between interactions and quantum geometry can drive weakly dispersive bosons into different exotic many-body phases. In this work we study a quasi flat-band model in one dimension that exhibits an extended pair-superfluid phase in the all-flat-band limit. Introducing single-particle hopping leads to an intriguing competition with a more conventional single-particle superfluid: we find that the pair superfluid remains stable for a finite range of the hopping strength until the system eventually transitions into the conventional superfluid phase. In our study, we make use of a variational Gaussian state approach that provides a unified description of the single-particle and pair superfluid phases, regarding both the ground state wavefunction and the collective excitation spectrum. In particular, we derive a general relation between the speed of sound and a ``quantum geometric kernel'', thereby extending earlier connections to the quantum metric, which relied on single-particle mean-field theory. This approach is combined with insights from the two-boson problem and exact diagonalization to map out the full phase diagram of the model. Our results show that the Gaussian approach is a versatile tool for studying a broad range of superfluid phases of interacting bosons in multi-orbital lattices.
\end{abstract}

\maketitle


\section{Introduction and overview}

For weakly dispersive particles, interaction effects become dominant and give rise to strong complex correlations, often leading to unconventional collective phenomena. In particular, the fate of superfluidity in interacting many-body systems hosting flat single-particle bands has been the topic of several studies, addressing both bosonic~\cite{Huber2010, Takayoshi2013, Tovmasyan2013_pair_condensation, Julku2021, Julku2021_excitations_of_BEC, Salerno2023, Julku2023, Amelio2024Lasing,huhtinen2026stabilityflatbandboseeinsteincondensation} and superconducting fermionic systems~\cite{peotta2015, Herzog2022, huhtinen2022revisiting, thumin2023, Tian2023, Yu2025}. Remarkably, it has been shown that superfluidity can persist even in the absence of single-particle dispersion: a finite sound velocity and non vanishing superfluid stiffness may arise in flat-band systems, provided that the underlying Bloch wavefunctions possess a nontrivial geometric structure~\cite{peottahuhtinentorma2025}.

\begin{figure*}[t]
    \centering
\includegraphics[width=0.9\linewidth]{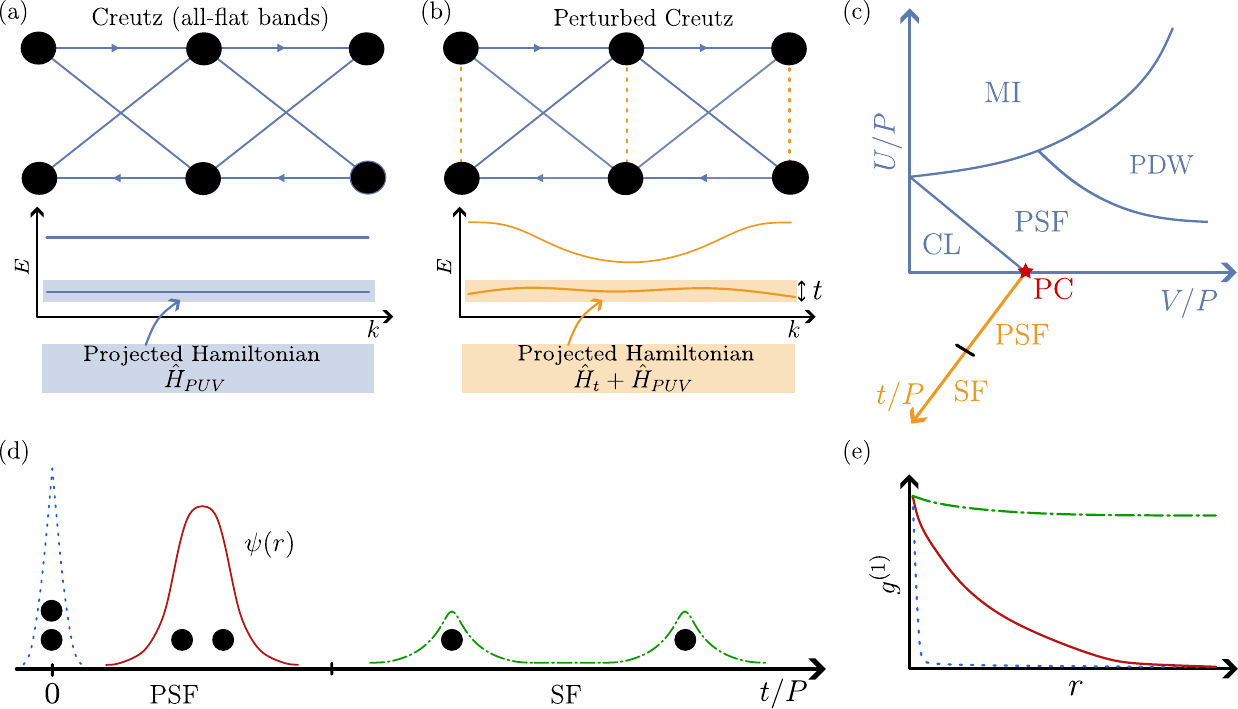}
    \caption{
    Sketch of the scope of the paper. 
    (a) The Creutz ladder is an all-flat band model; when we consider bosons with an interaction small compared to the band gap, we can project the low energy physics to the lowest band. The projected Hamiltonian is $\hat{H}_{PUV}$, comprising pair hopping processes (with rate $P$), and on-site and nearest-neighbor interactions (with strengths $U$ and $V$, respectively).
    (b) Perturbing the Creutz ladder by adding inter-leg hopping induces a single-particle kinetic term $\hat{H}_t$. A similar Hamiltonian holds for other projected quasi-flat band systems; in that case, the couplings $P,U,V$ (together with correlated hopping contributions) correspond to a low harmonic expansion of the quantum geometric kernel of Eq. (\ref{eq:QGK}).
    (c) In blue, the schematic phase diagram of the $\hat{H}_{PUV}$ Hamiltonian at unit filling. Mott insulating (MI), clustered (CL), pair density wave (PDW), pair superfluid (PSF) and pair condensate (PC) phases exist. A single-particle superfluid is instead forbidden by the local symmetry of $\hat{H}_{PUV}$, but can be stabilized above a critical $t$ (orange arrow). The Gaussian state approach works well in the PSF and SF regimes close to this line.
    (d) The PSF-SF competition can be understood at the 2-body level, as an unbinding transition occurring at large $t/P$. The radius of the bound state in the low density PSF regime determines the exponential decay length of the single-particle coherence function $\gsp$(r), as shown in panel (e). In particular, the blue dashed line corresponds to the {\it tightly bound pair} limit holding in the $\hat{H}_{PUV}$ case, where $\gsp(r>0)=0$ because of the local symmetry. The SF regime features instead quasi-long-range order (dash-dotted green line).
    }
    \label{fig:intro}
\end{figure*}

Bosonic flat and quasi-flat bands can be engineered in different physical platforms, exploiting geometrical frustration or destructive Aharonov-Bohm interferences~\cite{Leykam2018artificial}. This physics arises naturally in frustrated quantum magnets~\cite{Zhitomirsky2010,Riberolles2024}. Moreover,
Aharonov-Bohm cages arise in $\pi$-flux plaquettes, which are easily realizable with superconducting circuits~\cite{Martinez2023,Rosen2025}, whereas Floquet engineering allows for arbitrary fluxes~\cite{Alaeian2019,molinelli2026chiralbondorderedphasestriangularladder}. Optical lattices provide another powerful route to engineering flat bands for ultracold atoms, as done for the kagome~\cite{Jo2012} and Lieb lattices~\cite{Taie2015}. Regarding weakly interacting bosons, flat bands and Aharonov-Bohm caging have been observed in photonics~\cite{Leykam2018,Mukherjee2018}, while lasing in flat bands has been achieved in arrays of exciton-polariton micropillars~\cite{baboux2016,Harder2020} or in silicon metasurfaces~\cite{Eyvazi2025}. Finally, it was recently proposed that pair hopping processes can be tuned and made dominant over single-particle hopping in fluxonium quantum simulators~\cite{Amelio2026quantumsimulationfluxoniumqutrit} or Floquet-driven optical lattices~\cite{Goldman2023}.

At the theoretical level, previous studies have relied on single-particle mean-field theory, finding that, under some assumptions, the Bogoliubov speed of sound can be directly related to the quantum metric~\cite{Yu2025} of the flat band~\cite{Julku2021_excitations_of_BEC, Julku2021, Julku2023, Salerno2023, Amelio2024Lasing}. However, in all-flat band systems, the existence of local symmetries forbids 
any single-particle coherence~\cite{Tovmasyan2013_pair_condensation,Tovmasyan2018,Burgher2025},
while pair superfluidity can be stabilized~\cite{Tovmasyan2013_pair_condensation, Wang_2013, Tovmasyan2018, Luhmann2016, malakar2023}. In this regime, standard mean-field theory, based on the single-particle coherent state ansatz, is therefore conceptually inappropriate. Within the Gutzwiller ansatz, instead, it is possible to describe both single-particle and pair superfluid states, together with Mott insulators~\cite{Wang_2013, Luhmann2016, malakar2023}; however, in this approach, spatial correlations are not captured. Moreover, there have not been systematic comparisons between mean-field theory predictions and numerical calculations.

In this work, we combine exact analytical and numerical methods together with variational Gaussian states to understand the competition of pair and standard superfluidity in quasi-flat bands.
While a superfluid is typically defined by a finite speed of sound and 
(quasi-)long-range order of the single-particle coherence function $\gsp(r)$,
 pair superfluids feature coherence of pairs, quantified by  $\gp(r)$.
In practice, we will focus on the $tJPUV$ toy model, defined below through a low harmonic expansion of the quantum geometric kernel, an object which describes in full generality the interactions projected into a Bloch band.

As a first step, illustrated in Fig.~\ref{fig:intro}.(a), we consider the all-flat band case ($\hat{H}_{PUV}$ Hamiltonian). In this scenario, single-particle coherence is forbidden by a local symmetry. As sketched in blue in panel (c), the rich phase diagram of the $PUV$ model contains Mott insulator (MI), 
pair density wave (PDW), and clustered (CL) states, together with the pair superfluid (PSF) 
and an exact pair condensate (PC) point (which is, strictly speaking, not a pair superfluid).

Having identified the PSF region, the all-flat band case is perturbed by introducing single-particle
hopping ($\hat{H}_t$) and density-correlated single-particle hopping ($\hat{H}_J$), see Fig.~\ref{fig:intro}.(b). These terms break the local symmetry and drive the competition between the PSF and the single-particle superfluid (SF). Using exact diagonalization (ED) in one dimension, the PSF-SF transition is found at a finite value of the hopping rate. This physics can be intuitively understood by analytically obtaining the two-boson bound state: the unbinding of the pair determines the transition line in the low density limit (sketched in panel (e)), and its binding radius 
sets the exponential decay rate of the $\gsp$ coherence function in the PSF phase (panel (e)). Correlated hopping can also lead to collapse instabilities and atypical  correlation patterns.

Motivated by these analytical and numerical insights, we propose to generalize the bosonic Gaussian state approach~\cite{weedbrook2012,Guaita2019, Hackl2020} to 
describe both single-particle and pair condensation, and capture non-trivial two-body correlations. 
By comparing with ED,
we demonstrate that this ansatz provides 
a good theoretical understanding of the PSF versus SF competition and of the ground state correlations.
Moreover, we find a remarkable agreement between ED and the Gaussian method for the PSF collective excitation spectrum and dynamical structure factor, while standard mean-field theory yields qualitatively wrong results.
In general, the collective modes and the speed of sound are
determined by the Bloch wavefunctions of the band, which enter the equations in a complex way; nonetheless, in the all-flat band ($\hat{H}_{PUV}$) limit, it is possible to 
provide an explicit expression of the speed of sound in terms of 
the low harmonics of the quantum geometric kernel.

The paper is structured as follows. In Sec.~\ref{sec:generalmodel}, we define the quantum geometric kernel and introduce the $tJPUV$ model, which describes the quasi-flat band phenomenology in quite general terms. In Sec.~\ref{sec:ED}, we combine exact analytical insights and ED to obtain the phase diagram of the $tPUV$ model ($\hat{H}_{tPUV}=\hat{H}_{t}+\hat{H}_{PUV}$). The Gaussian state approach is developed in Sec.~\ref{sec:gauss}, where we investigate ground state correlations and calculate the dynamic structure factor. Conclusions and perspectives are drawn in Sec.~\ref{sec:conclusions}.
The impact of $\hat{H}_{J}$ is studied in Appendix \ref{app:correlated_hopping}, where analogies and differences with respect to $\hat{H}_t$ are discussed. In Appendix \ref{app:PairCondensate} we prove that the pair condensate is the exact ground state at a special point in the phase diagram. In Appendix \ref{app:Bogo} we give the details concerning the relation between the Bogoliubov speed of sound and the quantum geometry of the band. Finally, in Appendix \ref{app:dynamical_matrix} we give the technical details about the computations of the collective excitations within the variational Gaussian formalism.

\section{General framework and model}
\label{sec:generalmodel}

In this paper, we study interacting bosonic systems in translationally invariant multi-orbital lattices. For simplicity, we will assume that the low-energy physics occurs  within the lowest Bloch band of the system, which is well isolated in energy by a large gap, so that band mixing induced by the boson-boson interactions is negligible.
The projected Hamiltonian of the system can be expressed as 
\begin{equation}
    \hat{H} = 
     \sum_{\mathbf{k}} \epsilon_{\mathbf{k}}
    \hat{b}^\dagger_{\mathbf{k}} \hat{b}_{\mathbf{k}}
    +
      \sum_{\mathbf{k}\mathbf{p}\mathbf{q}} 
    h(\mathbf{k},\mathbf{p},\mathbf{q})
    \hat{b}^\dagger_{\mathbf{k}+\mathbf{q}} \hat{b}^\dagger_{\mathbf{p}-\mathbf{q}} \hat{b}_{\mathbf{k}} \hat{b}_{\mathbf{p}}.
    \label{eq:general_projected_H}
\end{equation}
where $\hat{b}^\dagger_{\mathbf{k}}$ creates a boson with Bloch momentum $\mathbf{k}$
and $\epsilon_{\mathbf{k}}$ is the single-particle dispersion of the lowest band. We have also introduced the ``quantum geometric kernel''~\cite{Amelio2024Lasing}
\begin{equation}
    h(\mathbf{k},\mathbf{p},\mathbf{q})
    = 
    \frac{1}{2 
     } 
    \sum_{\alpha\beta}
    U_{\alpha\beta}(\mathbf{q})
    u^*_{\mathbf{k}+\mathbf{q}}(\alpha) 
    u^*_{\mathbf{p}-\mathbf{q}}(\beta)
    u_{\mathbf{k}}(\alpha)
    u_{\mathbf{p}}(\beta),
    \label{eq:QGK}
\end{equation}
which encodes the information about the Bloch functions 
$u_{\mathbf{k}}(\alpha)$ 
determining the projection of the bare interaction $U_{\alpha\beta}(\mathbf{q})$; here, $\alpha
$ labels the orbital of each unit cell. 
 We note that,
 when $U_{\alpha\beta}=2$,
 the quantity
$
\sqrt{1 -  h(-\mathbf{k},\mathbf{k},2\mathbf{k})}
$, obtained from the quantum geometric kernel,
corresponds to the Hilbert-Schmidt distance between the states at momenta $\pm\mathbf{k}$.

In the following, we will focus on the {\em quasi-flat band} regime, where the $h$ term is large compared to the single-particle dispersion $\epsilon$, the {\em flat band} limit corresponding to $\partial_{\mathbf{k}} \epsilon_{\mathbf{k}}=0$. In these regimes,
the interactions and the geometry of the Bloch functions play a dominant role.

It is often insightful to recast the Hamiltonian of Eq.~\ref{eq:general_projected_H} into real space by introducing the Wannier operators $\hat{b}^\dagger_{\mathbf{R}}$. A non-trivial Bloch geometry manifests itself in the presence of correlated hopping and pair hopping terms, as well as nonlocal interactions, even if the original model features only a simple on-site Hubbard interaction
$U_{\alpha\beta}(\mathbf{q}) = U_0 \delta_{\alpha\beta}$. These extended correlated hopping and interaction terms can have a range of many unit cells; however, the main qualitative features can be understood by considering a few nearest neighbor terms, or equivalently, the low harmonics of $h(\mathbf{k},\mathbf{p},\mathbf{q})$.

\subsection{The $tJPUV$ model}

Following this philosophy, we now introduce the one-dimensional $tJPUV$ model, written as the sum of three contributions
\begin{equation}
    \hat{H}_{tJPUV} = \hat{H}_t +\hat{H}_J  + \hat{H}_{PUV} \;.
    \label{eq:tJPUV_H}
\end{equation}
The last term corresponds to an all-flat band system such as the Creutz ladder~\cite{Takayoshi2013,Tovmasyan2013_pair_condensation}, see Fig.~\ref{fig:intro}.(a). In this case, the existence of a local symmetry only allows for pair hopping, on-site interactions, and nearest neighbor interactions, respectively. A flat band in the presence of dispersive higher bands supports a correlated hopping term with coupling $J$, which is captured by the second term. Finally, we also incorporate single particle dispersion, captured by the first term, as illustrated in Fig.~\ref{fig:intro}.(b); for the case of the Creutz ladder, this single-particle hopping term is introduced by the dashed hopping links.

Let us now go over the different terms in more detail. The first term describes standard single-particle hopping processes,
\begin{equation}
    \hat{H}_t = 
    -t \sum_j \hat{b}_{j+1}^\dagger
    \hat{b}_j + h.c.,
\end{equation}
which give rise to a finite bandwidth of the single-particle dispersion, so that in this context
a nonzero $t$ arises as a perturbation from the flat band limit. The site index $j$ runs over $j=1,\dots,L$, where a chain with periodic boundary conditions and length $L$ is considered.
In standard superfluids, the kinetic term $\hat{H}_t$ is dominant and leads to
large single-particle coherences, quantified by the coherence function
\begin{equation}
\gsp(r) = \frac{1}{nL} \sum_j \langle  \hat{b}_j^\dagger \hat{b}_{j+r} \rangle,
\end{equation}
with $n$ being the boson density.
Note that in this 1D context, superfluidity refers to quasi-long range order and an algebraically decaying $\gsp(r)$, with an exponent predicted from the theory of Luttinger liquids~\cite{Giamarchi2003}.

The second term in our model is responsible for {\em correlated} single-particle hopping,
\begin{equation}
    \hat{H}_J = -J \sum_j \hat{b}_{j+1}^\dagger \hat{n}_j \hat{b}_{j-1} + h.c.,
    \label{eq:HJ}
\end{equation}
where $\hat{n}_j = \hat{b}_j^{\dagger}\hat{b}_j$.
At the single particle level, this term  does not induce any dispersion and indeed can arise
in flat-band lattices. For example, in the sawtooth chain,
$\hat{H}_J$ is present together with other longer-range correlated single-particle hopping processes which lead to similar dynamics~\cite{Huber2010}.
These terms lead to the single-particle superfluidity observed in the sawtooth chain.
Since the treatment of the $\hat{H}_J$ term  parallels the one for $\hat{H}_t$ (up to the relevant differences, explained where appropriate),
we have collected the results
concerning $\hat{H}_J$ in Appendix \ref{app:correlated_hopping}, so to streamline the presentation of the paper.

Both $\hat{H}_t$ and $\hat{H}_J$
possess standard symmetries, namely
number conservation, translational invariance, inversion and time-reversal symmetry. (This particular choice of correlated hopping $\hat{H}_J$
actually conserves the total particle number of even and odd sites separately.)
In contrast, the third contribution
\begin{multline}
    \hat{H}_{PUV} = 
    \sum_j  \left[ - P (\hat{b}^{\dagger 2}_{j} \hat{b}_{j+1}^2 + \text{h.c.})  +
    \right.
    \\
    \left.
    U \hat{n}_j (\hat{n}_j-1) + V \hat{n}_j \hat{n}_{j+1} \right],
    \label{eq:PUV_Hamiltonian}
    \end{multline}
also possesses an extensive set of {\em local symmetries}.
The conserved quantity in this case is the parity of the number of bosons on {\em each} site
$e^{i\pi \hat{n}_j}$~\cite{Tovmasyan2018}.
The parity operator indeed commutes  with each individual pair hopping and interaction term.

The pair hopping term allows for the transport of particles in pairs;
on the other hand, single-particle
superfluidity is prevented by Elitzur's theorem, forbidding spontaneous symmetry breaking for local gauge symmetries~\cite{Elitzur1975,Burgher2025}.
As a consequence, $\hat{H}_{PUV}$ supports
a pair superfluidity (quasi-)long range order, captured by the 
pair coherence function
\begin{equation} \label{eq:g_pair}
    g^{(1)}_{\rm pair}(r) = \frac{1}{nL} \sum_j \langle (\hat{b}_j^\dagger)^2 \hat{b}^2_{j+r}
\rangle.
\end{equation}
The ratios $t/P$ and $J/P$ then determine the competition between single-particle and pair superfluids, as we will see in the following.

The connection with Eq. (\ref{eq:general_projected_H}) is stressed by rewriting the Hamiltonian in momentum space:
\begin{equation}
   \hat{H}_{tJPUV} 
   =
   \sum_k
   \varepsilon_k 
   \hat{b}_{k}^\dagger
   \hat{b}_{k}
   +
   \sum_{kpq}
   h(k,p,q) \hat{b}_{k+q}^\dagger
   \hat{b}_{p-q}^\dagger
   \hat{b}_{p}
   \hat{b}_{k}
   \label{eq:hJPUV}
\end{equation}
with
$\varepsilon_k
= -2 t \cos k - \mu 
$ being the single-particle dispersion,
$\mu$ the chemical potential,
and 
\begin{equation} \begin{aligned}[b]
h(k,p,q) &=  h(|k+p|,|q|) \\
&= \frac{1}{L} \Big( U  + V \cos (q) - 2P \cos (k+p) \\
& \qquad\qquad\qquad  - 2J  \cos(k+p)\cos(q)  \Big),
\end{aligned} \end{equation}
embodying the quantum geometry of the bands. In other words, $J,P,U$ and $V$ correspond to the first harmonics of the quantum geometric kernel.

\begin{figure*}[t]
\includegraphics[width=1.0\linewidth]{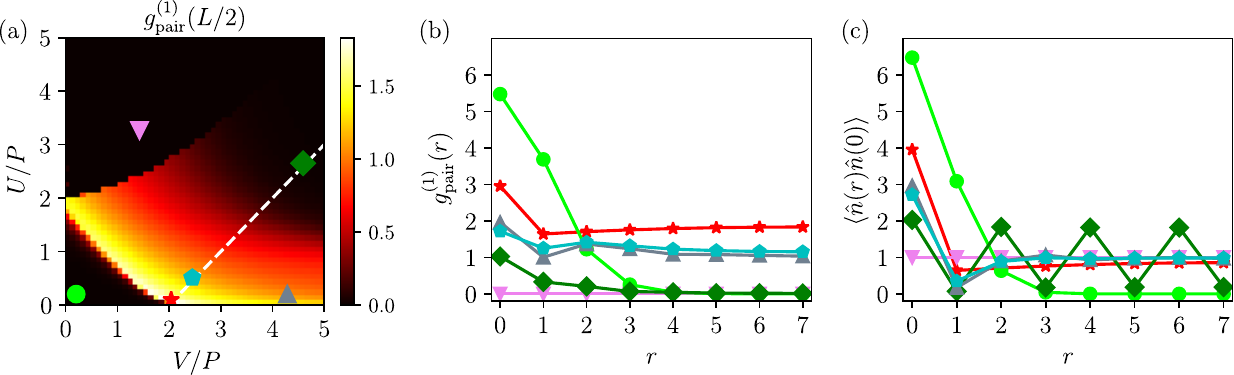}
\caption{Phase diagram of $H_{PUV}$ obtained via ED for a ring of $L=14$ sites at unit filling as a function of the couplings $V/P$ and $U/P$. Panel (a) displays the pair coherence function $\gp(r)$ [Eq.~\ref{eq:g_pair}] at the maximal distance $r=L/2$, while panel (b) and (c) show the behavior of the pair coherence and density-density correlator as a function of distance $r$, for a few selected points. The parameters corresponding to the selected points are indicated in panel (a), and correspond to the six following regimes: pair superfluid (cyan pentagon), clustered state (green circle), Mott insulator (pink downward triangle), pair density wave (green square), a mixture of pairs and heavy piles of pairs (grey upward triangle), and an ideal pair condensate (red star).}
\label{fig:correls}
\end{figure*}

\subsection{An example:  $\hat{H}_{tPUV}$ from the  Creutz ladder}
\label{ssec:Creutz}

Before proceeding with the  analysis of ground states and collective excitations,
we would like to provide a concrete example
of a lattice model
where the low-energy  physics is described, upon projection, by $\hat{H}_{tPUV} = \hat{H}_{t} + \hat{H}_{PUV}$
and $P,U,V$ can be conveniently tuned.
A necessary condition for the extensive local symmetry underlying $\hat{H}_{PUV}$ 
is that the single-particle spectrum consists only of all-flat bands~\cite{Tovmasyan2018}.
Here, we consider a paradigmatic model of all-flat band physics, the Creutz ladder~\cite{Creutz1999,Takayoshi2013,Tovmasyan2013_pair_condensation,Junemann2017}. Previous works focused on on-site Bose-Hubbard  interactions and 
explored
pair-superfluidity~\cite{Takayoshi2013,Tovmasyan2013_pair_condensation}, many-body localization phenomena~\cite{Kuno2020,Zhang2023,Burgher2025}, topological edges and few-body physics~\cite{Zurita2019,Pelegri2024}.
Here, we also consider the contribution of intra-rung interactions, so that the Hamiltonian reads
\begin{equation} \label{eq:H_Creutz}
\begin{aligned}[b]
\hat{H}_{\rm C} = -J_0 & \sum_j \Big[ i\hat{a}_{+,j+1}^\dagger \hat{a}_{+,j}  
         - i\hat{a}_{-,j+1}^\dagger \hat{a}_{-,j} \\
          & \qquad + \hat{a}_{+,j+1}^\dagger \hat{a}_{-,j} 
         + \hat{a}_{-,j+1}^\dagger \hat{a}_{+,j}
         + h.c. \Big] \\
        + & \sum_j \frac{U_0}{2} (n_{+,j}^2 + n_{-,j}^2) 
       + 
        \sum_j U_1 n_{+,j} n_{-,j} \;.
\end{aligned}
\end{equation}
Here, $\hat{a}^\dagger_{+,j}$ and 
$\hat{a}^\dagger_{-,j}$  respectively create particles on the $+$ and $-$ orbital of the $j$-th unit cell.
Moreover, we include both on-site and intra-rung interactions, with strengths $U_0$ and $U_1$, respectively.

The kinetic part of $ \hat{H}_{\rm Creutz}$
supports two flat bands at energies $\pm 2J_0$.
One eigenbasis is provided by compactly localized Wannier orbitals \cite{CL_Wannier_Sathe2021}; for the lowest band, the Wannier creation operators are given by 
\begin{equation}
    \hat{b}^\dagger_j = \frac{1}{2}\left(
    \hat{a}_{+,j}^\dagger 
    + i\hat{a}_{+,j+1}^\dagger
    + i\hat{a}_{-,j}^\dagger
    + \hat{a}_{-,j+1}^\dagger
    \right).
    \label{eq:Wannier}
\end{equation}
If $J_0 \gg U_0,U_1$, one can project out the upper band, to obtain
\begin{multline}
   \hat{H}_{\rm Creutz}^{\rm proj}
   =
   \frac{1}{16}\sum_j \bigg[ - (U_0 - U_1)(b_j^{\dagger 2} b_{j+1}^2 + h.c.) +
   \\
   +
    2(U_0 + U_1) \rho_j^2 + 4U_0 \rho_j \rho_{j+1}
    \bigg].
    \label{eq:H_Creutz_proj}
\end{multline}
This has exactly the same form as Eq.~(\ref{eq:PUV_Hamiltonian}), where,
by varying the ratio $U_0/U_1$, one can span the line
$V=U+2P$, represented by the white dashes in the $(U/P, V/P)$ plane of
Fig.~\ref{fig:correls}.(a). As we will see below, this allows one to access different interesting phases of the system.
In contrast, in previous studies only on-site interactions were considered (i.e. $U_1=0$, resulting in $V/P=4, U/P=2$). As a consequence, to promote the PSF over insulating phases, one needed to operate in a small density range~\cite{Takayoshi2013,Tovmasyan2013_pair_condensation}.

Finally,
we consider the  inter-leg hopping term   
$- 2t  \sum_j  (\hat{a}_{+,j}^\dagger \hat{a}_{-,j}  + h.c. )$,
indicated by the dashed orange links in Fig.~\ref{fig:intro}.(b).
As shown in Ref.~\cite{Tovmasyan2013_pair_condensation},
for $t \ll J_0$,
this perturbation, once rewritten in the Wannier basis of Eq. (\ref{eq:Wannier}),   induces 
the finite single particle dispersion term
$\hat{H}_t$
 without perturbing the eigenstates at the leading order, hence preserving $U,V,P$.
 Notice that  
 $t$ is
 possibly of the same order as $U,V,P$
 and independently tunable.
The $\hat{H}_J$ term, instead, is not present for the Creutz ladder, because it is not compatible with the local symmetry; $\hat{H}_J$-like terms would naturally arise in other not-all-flat band systems (as  the sawtooth chain or the kagome lattice).

\section{Phase diagram from Exact Diagonalization}
\label{sec:ED}

In this section, we will map out the ground-state phase diagram of the Hamiltonian of Eq.~\ref{eq:tJPUV_H} using exact diagonalization (ED), supported by analytical arguments holding in specific limits. The phase diagram is very rich and contains phases (such as Mott insulators and pair density waves) whose correlations cannot be captured in a Gaussian state treatment. For these reasons, the main goal of this section is to identify the range of parameters where superfluidity and pair superfluidity are expected, so to focus the Gaussian state analysis in Sec.~\ref{sec:gauss} onto this regime.

We used ED to calculate the ground state of the $tJPUV$ model and evaluated the correlation functions that allow us to discriminate between the different phases. Our ED implementation relies on the QuSpin library~\cite{Weinberg2017,Weinberg2019}. We start by considering the all-flat-band limit ($t=J=0$) limit (Sec.~\ref{ssec:pd_all_flat}), and afterwards study the effect of adding the single-particle hopping term (Sec.~\ref{ssec:pd_t}). The effect of the correlated hopping term is analyzed in Appendix \ref{app:correlated_hopping}. Finally, in Sec.~\ref{ssec:2body} we treat the limiting case of a two-boson system analytically.

\subsection{Phase diagram of $\hat{H}_{PUV}$}
\label{ssec:pd_all_flat}

In the all-flat-band limit $t=J=0$, the single-particle coherence function is identically zero, i.e. $\gsp(r)=0, \forall r\neq 0$ and standard superfluidity is forbidden by Elitzur's theorem. On the contrary, the pair correlation function plays a central role: In the pair superfluid phase, $\gp(r)$ decays algebraically with $r$, for the pair condensate true long range order is expected, while insulating phases feature an exponential decay. As a diagnostic for the insulating phases we also compute the density-density correlation function $\langle \hat{n}_j \hat{n}_{j+r} \rangle$.

In Fig.~\ref{fig:correls}, we present our results for the phase diagram for unit filling ($n = 1$) as a function of the couplings $U/P$ and $V/P$ : in panel (a) we plot $g^{(1)}_{\rm pair}(L/2)$, as a proxy of the decay rate of the coherence in space, whereas in panel (b) and (c) we plot the full pair correlation function and the density-density correlation function for a few representative points. 
 While for large $U/P$ and $V/P$ insulating states appear, we find an extended region in proximity of  $U/P=0, V/P=2$ where a pair superfluid phase is stabilized; the finite value of $V/P$ prevents collapse instabilities. 
In total, we identify six different regimes, which we now describe.

\emph{Pair superfluid phase (cyan pentagon).} 
The cyan pentagon denotes a pair superfluid state, characterized by a barely visible decay of $\gp(r)$. The pair  coherence is maximal in  the vicinity of the pair condensate point $U=0, V=2P$.

\emph{Cluster state (green circle).} %
At small repulsive interactions the bosons tend to collapse. The line $U+V=2P$, derived setting to zero the mean-field 
interaction energy
of the fluid, provides an approximate boundary for this instability. This collapse is revealed by the correlation functions that reflect the bosons being clustered in a single droplet. The clustering is driven by the pair hopping term: this is both a kinetic and a two-body interaction contribution, which is enhanced at high density.

\emph{Mott insulator phase (pink triangle).} 
When $U/P$ is the dominant scale, the pure Mott-insulating state 
\begin{equation}
|\Psi\rangle_{\text{Mott}} = \bigotimes_j \hat{b}^\dagger_j |0\rangle
\end{equation} 
is the ground state. Importantly, this is an exact ground state, on which the pair hopping operator gives zero. From a symmetry point of view, this is indeed the only state of the Hilbert space at filling 1 to possess an odd parity of bosons on each site. The transition to 
$\ket{\Psi}_{\text{Mott}}$ 
is therefore sharp, occurring with a jump of correlation functions. Indeed, we find $\gp(r)=0$ and $\langle \hat{n}_j \hat{n}_{j+r} \rangle=1$ in the Mott insulating phase.

\emph{Pair density wave phase (green square).} 
When $V/P$ dominates, instead, a staggered pair density wave is stabilized. In the $P \to 0$ limit and $V>U$, the ground state takes the form 
\begin{equation}
    |\Psi\rangle_{\text{PDW}} = \frac{1}{\sqrt{2}} \left[ \bigotimes_j \frac{(\hat{b}^\dagger_{2j})^2}{\sqrt{2}} + \bigotimes_j \frac{(\hat{b}^\dagger_{2j+1})^2}{\sqrt{2}} \right] |0\rangle
\end{equation}
In this phase, as visible from the green squares, $\gp(r)$ decays exponentially, while $\langle \hat{n}_j \hat{n}_{j+r} \rangle$ displays a marked staggered pattern. 
We remark that, to our understanding, proper supersolidity is not expected at commensurate fillings, in neither 1D~\cite{Kuhner2000} or 2D~\cite{Sengupta2005} lattices.

\emph{Regime with a mixture of pairs and heavy piles of pairs (gray triangle).} %
Instead, for $V/P \gg 1$ and $U/P \ll 1$,  the most efficient way to avoid the cost of having particles in neighboring sites while allowing for pair hopping processes, is to stock more pairs on the same site (which is possible by virtue of the small $U$), in order to leave space for a fraction of the pairs to move.
Overall, the correlators in this regime display the pair coherence typical of pair superfluids, but manifest a stronger tendency towards bunching at $r=0$ and towards anti-bunching at $r=1$.

\emph{Pair condensate (red star)}. %
Finally, the red star denotes the pair condensate state arising at the point with $V/P=2$ and $U=0$. It turns out that the {\em pair condensate} state
\begin{equation}
    |\Psi\rangle_{\text{PC}} \propto \left( \hat{\Pi}_0^\dagger\right)^{N/2}
    |0\rangle,
\end{equation}
with the pair creation operator of total momentum $Q$
\begin{equation}
    \hat{\Pi}_Q^\dagger =  \sum_j e^{iQj}(\hat{b}_j^\dagger)^2
\end{equation}
is the exact ground state of $\hat{H}_{PUV}$ for $V=2P>0$ and $U=0$, in the $N$ particle sector (with $N$ even). This state 
$|\Psi\rangle_{\text{PC}}$
contains $N/2$ zero momentum pairs, and is the pair-condensate equivalent of a pure singe-particle condensate state. In complete analogy to non-interacting single-particle condensates, the system is actually not superfluid for this precise parameter set, as it turns out that the excitation spectrum is gapless and parabolic. We prove these statements in Appendix \ref{app:PairCondensate}.

Note that the white dashes, which denote the $V=2P+U$ line spanned by the Creutz ladder model upon varying $U_0/U_1$, encompass the pair condensate state, the pair superfluid regime and the pair density wave phase.
Moreover, we expect the insulating states to be suppressed at incommensurate densities, and the PSF region to have a larger extent.

\subsection{Phase diagram of $\hat{H}_{tPUV}$}
\label{ssec:pd_t}

\begin{figure*}[t]
\includegraphics[width=0.9\linewidth]{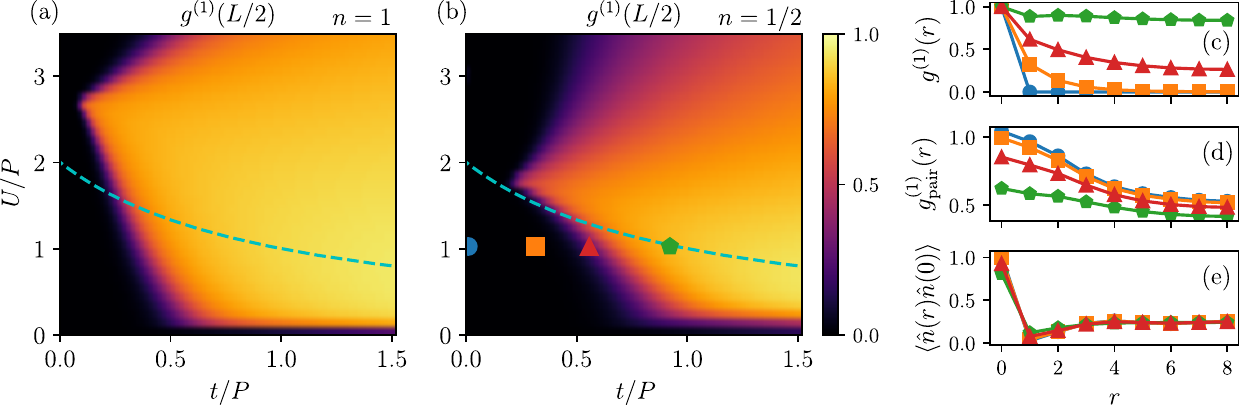}
\caption{
The effect of the single-particle hopping $t$ and the onsite interaction strength $U$ on the coherence of the system is investigated. We fix $V=2P$ and $J=0$. In panel (a), we display $g^{(1)}(L/2)$ at unit filling $L=N=14$, while half filling $L=16,N=8$ is shown in panel (b). The cyan dashed line corresponds to the 2-body result of Eq. (\ref{eq:2body_tUV_line}). In panels (c,d,e), we respectively report the single-particle coherence function, the pair coherences and density-density correlator as a function of distance $r$, for the four different points indicated in panel (b). The PSF at $t=0$ is indicated by the blue circle, the PSF at finite $t$ by the orange square, the SF phase by green pentagons, while the red triangle is taken in the vicinity of the SF-PSF transition.}
\label{fig:t_ED}
\end{figure*}

Let us now introduce the single particle hopping term, and investigate the stability of the pair superfluid phase that we identified in the previous section.  In Fig.~\ref{fig:t_ED}, we show the ED results for the ground state properties of $\hat{H}_{tPUV}$ as a function of $t/P$ and $U/P$, for fixed $V/P=2$ and $J=0$. 

In panel (a) we plot the single particle coherence at the largest accessible distance $\gsp(L/2)$, calculated  at unit filling. It is found that the pair superfluid phase is stable for small $t/P$ and is characterized by a single particle coherence that is exponentially small, while at sufficiently large $t$ a single-particle superfluid state emerges, with sizable single particle coherence.
Initially, we find that the critical $t/P$ decreases with $U/P$. This is expected since double occupations are penalized by a local repulsion, so that $U$ drives the pair superfluid into the single-particle superfluid phase. This behaviour is qualitatively consistent with the two-body unbinding transition discussed below in Eq. (\ref{eq:2body_tUV_line}), denoted by the cyan dashed lines. At large $U$,  superfluidity is instead suppressed through the competition with the Mott insulator phase.

In panel (b) we plot the results for the case of half filling. We find that the two-body prediction (cyan dashed line) becomes more accurate and, for small $U$, the critical $t/P$ is larger for smaller densities. 

Four points at $U/P=1$ are further analyzed in panels (c-e) through their single-particle and pair coherence function, as well as the density-density correlation function. The blue circle corresponds to the pair superfluid at $t=0$, where the single-particle coherence function is exactly zero as bosons always appear in pairs because of Elitzur's theorem, and the pair coherence function decays algebraically. The orange square corresponds to the situation at small $t$ in the pair-superfluid phase, where we see that the single-particle coherence function decays exponentially; this exponential decay can be understood in the low density regime from the wavefunction of the pairs being exponentially localized, with a binding radius increasing with $t$. The green pentagon is situated in the single-particle superfluid phase, for which we observe a slow algebraic decay of the single-particle coherence function and a decrease of the pair coherence function. Finally, the red triangle shows the crossover between the two superfluid phases. The density-density correlation function displays a very similar behavior in the four cases; in particular, the peak at $r=0$ reveals a large probability of finding preformed pairs also in the SF phase.

\subsection{Two-body physics}
\label{ssec:2body}

As anticipated, useful analytical insights can be obtained from the study of 
the two-boson ground state.
Considering an infinite chain, the state of the two bosons at zero total momentum can be written as
\begin{equation}
    |\Psi_{N=2}\rangle
    =
    \left[ 
    \sum_{r>0}
    \psi(r) 
    \sum_{j}
    \hat{b}^\dagger_{j} \hat{b}^\dagger_{j+r}
    +
    \frac{\psi(0)}{\sqrt{2}}  
     \sum_{j}
    \left( \hat{b}^\dagger_{j} \right)^2
    \right],
\end{equation}
where $\psi(r)$ represents the relative wavefunction.
The stationary Schr\"odinger equation reads
\begin{equation}
\begin{aligned}
    E \psi(0) &= \tilde{U} \psi(0) -2 \sqrt{2} t \ \psi(1) \\
    E \psi(1) &=  \tilde{V}\psi(1) -2 \sqrt{2} t \ \psi(0) - 2t \psi(2) \\
    E \psi(r) &=  -2  t \, \Big[\psi(r-1) + \psi(r+1) \Big], \quad \forall r \geq 2
\end{aligned}
\end{equation}
where we defined $\tilde{U} = 2U - 4P$ and $\tilde{V} = V - 2J$. Notice that these are the equations of a particle with hopping rate $2t$ (the two corresponding to the reduced mass) scattering against a potential $\tilde{U} \delta_{j,0} + \tilde{V} \delta_{j,\pm 1}$. In this comparison, the $\sqrt{2}$ factors comes from $\psi(r>0)$ being the symmetric superposition of the effective particle wavefunction at positive and negative  positions.

These equations are solved for a bound state using 
$\psi(r\geq 2) = e^{-\kappa r}$, with
$E = -4t \cosh(\kappa)$.
A self-consistent equation for the energy is found
\begin{equation}
    E
    =
    \frac{8t^2}{E - \tilde{V} - \frac{8t^2}{E - \tilde{U}}} + \sqrt{E^2 - 16t^2} \ .
    \label{eq:E2body}
\end{equation}
The edge of the continuum is given by $E=-4t$ and the binding energy of the pair is then defined from $E = -4t - |E_B|$.
In the low density limit, if collapse instabilities are not present, the pair superfluid to single-particle superfluid transition will then be approximated by the 
line $E=-4t$ (or $|E_B|=0$), at which the two-body state unbinds.
In terms of $\tilde{U}/t$ and $\tilde{V}/t$,
this occurs on the curve
\begin{equation}
    \frac{\tilde{U}}{2t}
    =
    -2 + \frac{2}{1+\tilde{V}/(2t)}
\label{eq:2body_tUV_line}
\end{equation}
which was plotted as a cyan dashed line in Figs. (\ref{fig:t_ED},\ref{fig:Gaussian_t}).
Notice that for $\tilde{V}=0$ any negative $\tilde{U}$ will lead to a bound state, as it is well known for an attractive potential in 1D and 2D; in the opposite limit $\tilde{V} \to +\infty$, one needs to have $\tilde{U}<-4t$ in order for the bound pair to be the ground state.
Also, a bound state exists for any $\tilde{V} <-2t$,
independently of the value of $\tilde{U}$.
Conversely, fixing $\tilde{U}$ and $\tilde{V}$, increasing $t$ will unbind the pairs at the critical hopping strength
$t_c = -\tilde{U}\tilde{V} / (2\tilde{U}+4\tilde{V})$.

\section{Gaussian state analysis}
\label{sec:gauss}

Previous studies of superfluidity in flat bands have relied on coherent states, assuming condensation at the single particle level. Then, the Bogoliubov treatment yields a superfluid speed of sound linked to the  Bloch wavefunctions of the flat band~\cite{Julku2021_excitations_of_BEC, Julku2021, Julku2023, Amelio2024Lasing}. However, in the previous section, we have demonstrated the existence of a pair superfluid phase, characterized by large pair coherences, and zero or exponentially decaying single particle correlations. A transition to a single particle superfluid occurs for sufficiently dispersive bosons or for a sizable correlated hopping rate. This physics is qualitatively captured by the binding of two bosons.
It is therefore compelling to analyze the PSF state and the transition to the SF beyond standard mean-field theory. In this section, therefore, we will use Gaussian states~\cite{weedbrook2012,Guaita2019, Hackl2020} as a variational class of states for describing the low-energy behavior of the $tJPUV$ model, enabling us to describe both the pair superfluid and single particle fluid in a unified framework, as well as the two-body physics that describes the model accurately in the low density limit. 

In Sec.~\ref{ssec:GaussGS}, we will extend the formalism of variational Gaussian states, as laid out in Ref.~\onlinecite{Guaita2019}, towards the description of quasi-flat band models; we will focus on determining the ground-state properties as well as the low-lying dynamics. In Sec.~\ref{ssec:Gauss_tJPUV}, we then apply this formalism to address the competition between PSF and SF phases under the influence of a hopping term and to study their collective excitations by computing the dynamical structure factor. Finally, the limiting scenario where all bosons occur in tightly bound pairs is analyzed in Sec.~\ref{ssec:Simon}.

\subsection{Gaussian states for quasi-flat band models}
\label{ssec:GaussGS}

The most general bosonic Gaussian state can be expressed as the exponential of a free bosonic Hamiltonian, containing linear and quadratic bosonic operators. However, the symmetries of the problem allow for simplifications, and often the most convenient representation is in terms of a product of exponentials rather than a unique exponential.
We consider here translationally invariant Gaussian states represented by the wavefunction
\begin{equation}
    |\Psi_g(\vec{x}_g)\rangle
    =
    \hat{\mathcal{U}}_g(\vec{x}_g) 
    | 0 \rangle 
    =
    \hat{\mathcal{D}}_g(\beta) \hat{\mathcal{S}}_g( \{ \lambda_k \})
    | 0 \rangle. 
\end{equation}
Here, $\vec{x}_g$ collects the variational parameters $\beta$,  the single-particle condensate amplitude, and $\{ \lambda_k \}$, the squeezing strengths for each momentum
$k$.
This state is normalized, since $\hat{\mathcal{U}}_g(\vec{x})$
consists of two unitary operators, the displacement
operator
\begin{equation}
    \hat{\mathcal{D}}_g(\beta) = \exp \left( \beta \hat{b}_0^\dagger - \beta^* \hat{b}_0 \right),
\end{equation}
where $\hat{b}_0^\dagger$
creates a boson at zero momentum,
and the squeezing operator
\begin{equation}
    \hat{\mathcal{S}}_g( \{ \lambda_k \}) = \exp \left( 
    \frac{1}{2}
    \sum_k 
    \lambda_k \hat{b}_k^\dagger
    \hat{b}_{-k}^\dagger
    - 
    \lambda^*_k \hat{b}_k
    \hat{b}_{-k}
    \right),
\end{equation}
which introduces 2-body correlations.

We consider the set of Gaussian states as a variational class parametrized by $\vec{x}_g$, so that the best approximation of the ground state is found by minimizing the variational energy
\begin{equation}
    E_g(\vec{x}_g) = \langle \Psi_g(\vec{x}_g) |\hat{H} |\Psi_g(\vec{x}_g)\rangle \;,
\end{equation}
where we consider the Hamiltonian of the form
\begin{equation}
    \hat{H} = \sum_k \varepsilon_k \hat{b}_{k}^\dagger \hat{b}_{k} + \sum_{kpq}
   h(k,p,q) \; \hat{b}_{k+q}^\dagger \hat{b}_{p-q}^\dagger \hat{b}_{p} \hat{b}_{k} \;.
\end{equation}
Combining the $U(1)$ number conservation symmetry and the fact the Hamiltonian contains only real parameters, the parameters $\beta$ and $\lambda_k$ can be chosen to be real. Moreover, we enforce $\lambda_k = \lambda_{-k}$ to remove the overparametrization arising from
the bosonic commutation of $ \hat{b}_k^\dagger \hat{b}_{-k}^\dagger$; this also explains the choice of the $1/2$ factor.
In the following, we also assume 
$h(k,p,q)=h(|k+p|,|q|)$ for the quantum geometric kernel, as in the case of the $tJPUV$ model; while this choice allows for a few arithmetic simplifications, all the results can be straightforwardly extended to a generic $h$.
 Then, the action of $\hat{\mathcal{U}}_g(\vec{x}_g)$ on the bosonic operators reads
\begin{equation}
\hat{\mathcal{U}}_g^\dagger 
\hat{b}_k
\hat{\mathcal{U}}_g
=
\beta \delta_{k,0} 
+
u_k \hat{b}_k
+
v_k \hat{b}_{-k}^\dagger,
\label{eq:Bogo_rotation}
\end{equation}
with $u_k = \cosh \lambda_k, v_k = \sinh \lambda_k$.

This rotation, together with Eq. (\ref{eq:hJPUV}), allows to calculate the energy expectation value
\begin{equation}
E_g(\vec{x}_g) = E_0 + \beta^2 E_2 + \beta^4 E_4,
    \label{eq:GS_energy}
\end{equation}
with 
\begin{align}
    E_0 &= \sum_k \left[ \varepsilon_k v^2_k + \Delta(k) u_k v_k + \frac{1}{2}\Xi(k)  v^2_k \right]\\
    E_2 &= \varepsilon_0 + 2 \Delta(0) + \Xi(0) \\
    E_4 &= h(0,0), 
\end{align}
where we  introduced
\begin{align} 
\Delta(k) &= \sum_p h(0,p-k) u_p v_p \\
\Xi(k) &=  2\sum_p \left[ h(k+p,0) + h(k+p,p-k) \right] v_p^2 \;.
\end{align}
This notation was chosen to highlight that $\Delta$ and $\Xi$ enter the equations in a similar way as the gap and self-energy in BCS theory.
We remark that, with our conventions, $\beta^2$ is extensive and scales  proportionally to $L$, while the $u_k$ and  $v_k$'s are intensive, and $h \sim 1/L$. As a result, $E_g$ is extensive. Then, differentiating by $\beta$ and $\lambda_k$ yields the saddle point equations
\begin{equation}
     \varepsilon_0 + 2 \beta^2 h(0,0)
     + 2 \Delta(0) +  \Xi(0) = 0,
\end{equation} 
\begin{equation}
    \frac{1}{2} \tanh 2\lambda_k
    = - \frac{\beta^2 h(0,k) + \Delta(k) }{\epsilon_k + 2 \beta^2 [h(k,0) + h(k,k)] +  \Xi(k)},
    \label{eq:tan2l}
\end{equation}
which we solve iteratively. We remark that these saddle point equations are expressed in terms of the quantum geometric kernel $h$ and relate the correlations characterizing the Gaussian ground state to the geometric properties of the Bloch bands via Eq. (\ref{eq:QGK}).

Besides ground-state properties, the Gaussian state framework also allows us to study the low-energy excitation spectrum~\cite{Guaita2019,Hackl2020}. The idea is to consider a slight perturbation of the Hamiltonian and use the Dirac-Frenkel time-dependent variational principle (TDVP) to capture the dynamics. Upon linearizing the TDVP equations and solving for the lowest energy modes, we find approximate excitations in the tangent space of the manifold of Gaussian states.

We use this linearized TDVP approach to investigate the dynamical structure factor. We use the density operators at momentum $Q$,
\begin{equation}
    \delta\hat{n}(Q)  = \frac{1}{L}\sum_k \hat{b}^\dagger_{k+Q} \hat{b}_k
- n \delta_{Q,0}
\end{equation}
which defines the dynamic structure factor
\begin{equation}
    S(Q,\omega) = 
     \int_{-\infty}^{+\infty} d\tau \ e^{i\omega \tau} \langle\delta\hat{n}(Q,\tau) \delta\hat{n}(-Q,0) \rangle.
\end{equation}
The recipe for computing $S(Q,\omega)$ numerically within the Gaussian state approach is detailed in Appendix \ref{app:dynamical_matrix}.

Note that, in Ref. \cite{Guaita2019}, single-particle superfluid states were addressed in the context of the Bose-Hubbard model with contact repulsive interactions, demonstrating quantitative corrections with respect to single-particle mean-field theory. In our treatment, we have provided an extension of the theory to general interaction kernels $h(k,p,q) = h(|k+p|,|q|)$. We will now show that this approach also describes the PSF regime, where qualitative differences with respect to mean-field theory are found.

\subsection{Results for the $tPUV$ model}
\label{ssec:Gauss_tJPUV}

While the theoretical framework explained in the previous Section applies to the generic Hamiltonian of Eq. (\ref{eq:general_projected_H}), under the only assumption of a translationally invariant ground state, we now use the Gaussian method to explicitly compute  the ground state and collective excitations of the $tPUV$ model. 
The effect of correlated hopping $\hat{H}_J$
is discussed in Appendix \ref{app:correlated_hopping}.

\begin{figure*}[t]
    \centering
\includegraphics[width=0.9\linewidth]{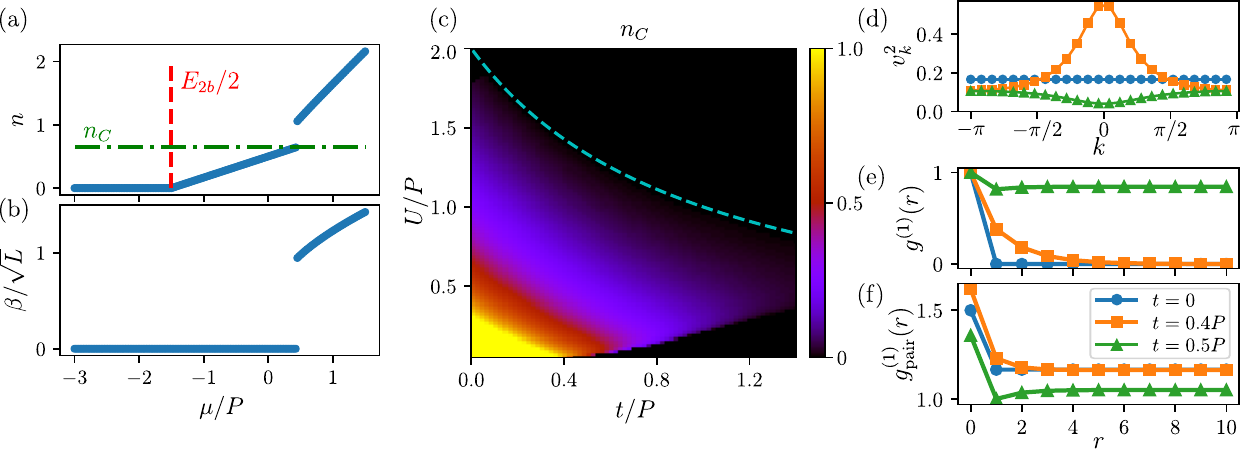}
\caption{ 
    Ground state within the Gaussian  approach.
    Panels (a) and (b) respectively display the density $n$ and condensate order parameter $\beta$ as a function of the chemical potential $\mu$, for fixed $t=J=0$ and $V=2P,U=P/2$. The density is non-zero above the two-body bound state threshold $\mu \geq E_{2b}/2$ (red dashed line).
    The PSF is stable till a critical density $n_C$ (green dashed-dotted line), above which single particle condensation occurs via a first-order phase transition.  
    In panel (c),
    the critical density $n_C$ is studied as a function of $U/P$ and $t/P$ for $V=2P$.
    The cyan dashed line corresponds to
    the 2-body binding transition from
    Eq.~(\ref{eq:2body_tUV_line}).
    The colorscale is saturated close to the origin.
    Panel (d) reports the Gaussian squeezing parameters $v_k$ for $t=0$ (blue), $t=0.4P$ (orange) and $t=0.5P$ (green), for a fixed $\mu=-P$.
    Panel (e) and (f) display the single particle and pair coherence correlation functions, for the same parameters.
    In all plots, the system size used for the numerical calculation was $L=50$.
    }
    \label{fig:Gaussian_t}
\end{figure*}

\subsubsection{Ground-state properties}

Since the Gaussian states are not eigenstates of the total particle number operator,
it is more convenient to adopt a grand-canonical approach and vary the chemical potential $\mu$ as a control parameter. We will then focus on the average density, 
\begin{equation}
n =\frac{\beta^2}{L} + \frac{1}{L} \sum_k v_k^2,    
\end{equation}
as a function of $\mu$. We show the typical behavior 
of $n(\mu)$
in Fig. \ref{fig:Gaussian_t}.(a), where we have fixed the parameters $(V/P,U/P)=(2,1/2)$ and no single-particle hopping $t=J=0$. In the lower panel, the condensate amplitude $\beta/\sqrt{L}$ is shown. In the pair superfluid phase with $\beta=0$, the density increases monotonically with $\mu$, until it reaches a critical value $n_C$ (green dash-dotted line). At this point, a first order phase transition occurs to a single-particle condensate phase, with a jump in both $n$ and $\beta$.

Thus, in spite of $t=J=0$, also the Gaussian state approach predicts a single-particle condensate at large $\mu$, in violation of Elitzur's theorem. While this is an artifact of the variational approach, it suggests that at large densities the system is susceptible to develop strong single-particle coherences as soon as $t$ is switched on. As a consequence, we expect that achieving the PSF phase in high density weakly interacting Bose fluids would be very difficult in practice, since any small disorder or non-ideality would 
perturb the flatness of the band and
result in a single-particle SF. For instance, this is the typical regime in experiments with photonic and exciton-polariton lattices.

Concerning the nature of the PSF to SF phase transition, one would be tempted to believe that it should be of second order and of the Ising type, given the breaking of $\mathbb{Z}_2$ symmetry. However, since the PSF hosts a Goldstone mode, the Coleman-Weinberg mechanism~\cite{Coleman1973} can 
turn the transition into a first-order one, as observed in renormalization group~\cite{Lee2004} and Quantum Monte Carlo~\cite{Ng2011} studies. Apart from the description of the phase diagram, the Gaussian state approach captures many interesting features of the PSF state, as we are going to see below.

Importantly, in the low density limit ($\beta \to 0, u_k \to 1, v_k \to 0$) the saddle point equation (\ref{eq:tan2l}) reduces to 
\begin{equation}
    \epsilon_k v_k + \sum_p h(0, k-p) v_p = 0,
\end{equation}
which is equivalent to the Schr\"odinger equation (times 1/2) for two bosons with zero total momentum, where the chemical potential needs to satisfy $\mu=E_{2b}/2$, $E_{2b}$ being the solution of the two-body problem in Eq. (\ref{eq:E2body}). This is indicated by the vertical red dashed line in Fig. \ref{fig:Gaussian_t}.(a). This argument is consistent with the intuition that Gaussian states capture two-body correlations and are exact in the two-body limit; in the low density limit, $v_k$
thus represents the relative wavefunction of each bound pair. This is further analyzed in panel (d),
where $v_k$ is plotted: for $t=0$ (blue line), the pairs are always perfectly bound in real space, resulting in a constant $v_k$; turning on $t$, makes $v_k$ peaked at zero momentum (orange), corresponding to less correlated particles in real space.   

These considerations are further analyzed in
Fig. \ref{fig:Gaussian_t}.(c),
where the critical PSF density $n_C$
is shown as a function of $U/P$
and $t/P$.
A large value of $n_C$ signifies that the PSF phase is robust against the formation of a single particle condensate when the chemical potential is increased.
The qualitative behavior resembles the ED results in Figs. \ref{fig:t_ED}.(a,b), where the PSF phase is suppressed by increasing either $U$ or $t$.
Moreover, in the low density regime, the 2-body threshold of 
Eq.~(\ref{eq:2body_tUV_line})
for the binding of a pair is exactly recovered, as indicated
by the cyan dashed line.
At small $U$ and large $t$, we find that the PSF is always unstable towards the formation of a single-particle condensate. 

Finally, the correlation functions can be computed using Eq. (\ref{eq:Bogo_rotation}),
yielding the single particle coherence function
\begin{equation}
    g^{(1)}(r)
    =
    \frac{\beta^2}{n L}
    +
    \frac{1}{n L} \sum_k
    e^{ikr}
    v_k^2,
\end{equation}
plotted in panel (e), and the pair coherence
\begin{multline}
    g^{(1)}_{\rm pair}(r)
    =
    \frac{\beta^4}{nL^2}
    +
    \frac{\beta^2}{nL^2} \sum_k 
    \left( 
    2 u_k v_k
    + 4
    e^{ikr}
    v_k^2
    \right)
    +
    \\
    +
    \frac{1}{nL^2} \sum_{k,p}
        \left[ 
    u_k v_k u_p v_p
    + 2
    e^{i(k+p)r}
    v_k^2 v_p^2
    \right],
\end{multline}
reported in panel (f). We have considered three representative points in the phase diagram: we fixed $(V/P,U/P)=(2,1/2)$ and $\mu/P=-1$, and took three values for the hopping strength $t/P=0$, $t/P=0.4$ and $t/P=0.5$. We clearly see three different regimes for the single particle coherence function: for $t=0$ it is identically zero for $r>0$, it decays exponentially for $t/P=0.4$, while it tends to a constant for $t/P=0.5$, for which the ground state has non-zero $\beta$. The pair coherences $g^{(1)}_{\rm pair}(r)$ are instead only weakly affected by the single particle hopping, displaying a slight decrease for increasing $t$. Overall, these results are in excellent qualitative agreement with the ED findings of Sec.\ref{sec:ED}.
As a final remark, we point out that the Gutzwiller ansatz was previously used to study 
superfluid, pair superfluid and Mott insulating phases in related models~\cite{Wang_2013, Luhmann2016, malakar2023}.
However, in the Gutzwiller treatment, spatial correlations are completely neglected and the low density limit is not correctly described.

\subsubsection{Collective excitations}

\begin{figure*}[t]
    \centering
\includegraphics[width=0.99\linewidth]{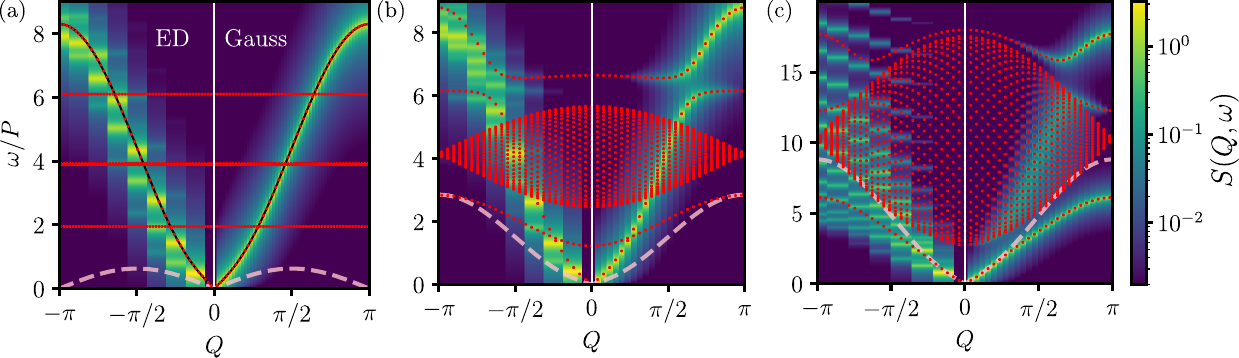}
    \caption{The dynamic structure factor $S(Q,\omega)$ was computed in both exact diagonalization (displayed for $Q \leq 0$) and the Gaussian state approach (in the $Q \geq 0$ half), for three different parameter regimes.
    The red dots denote the energies of the Gaussian collective modes, and the black line in panel (a) corresponds to the tightly bound pair calculation from Sec. \ref{ssec:Simon}.
    The pink dashed line is the Bogoliubov dispersion from Appendix \ref{app:Bogo}.
    The three panels correspond to the following regimes: 
    (a) tightly bound pair superfluid with $t=0, U=0.05P, V=2.05P,n=0.5$;
    (b)  pair superfluid with $t=0.2P, U=0.05P, V=2P,n=0.5$;
    (c)  normal superfluid with $t=0.8P, U=0.2P, V=2P,n=1.0$;
    }
    \label{fig:spectra}
\end{figure*}

We have evaluated the dynamic structure factor $S(Q,\omega)$ using both the Gaussian state approach 
and exact diagonalization. In ED, 
$S(Q,\omega)$ can be obtained 
efficiently by relying on Krylov space methods~\cite{senechal2010introductionquantumclustermethods}.
An artificial broadening of $\eta=0.1P$ has been chosen for the plots, and the colors encode a logarithmic scale.
The results for three different regimes 
($t=0$ PSF, $t>0$ PSF, SF)
are shown in Fig.~\ref{fig:spectra}, where for negative momenta we report the ED results, while the Gaussian state calculation is displayed for $Q>0$. 
The red points are the energies of the collective modes computed within the Gaussian state approach (corresponding to the eigenvalues of the dynamical matrix $M(Q)$ defined in Appendix \ref{app:dynamical_matrix}); some of these excited states are dark, in the sense that they have zero spectral weight.
As pink dashed lines,  we report the results of standard mean-field theory~\cite{Julku2021_excitations_of_BEC,Julku2021,Julku2023,Amelio2024Lasing}, where Bogoliubov linearization is performed on top of a single-particle condensate ansatz. The details  of this Bogoliubov dispersion are provided in Appendix  \ref{app:Bogo}.

In panel (a), we study the PSF regime at $t=J=0$ and $(V/P,U/P)=(2.05,0.05)$ and at half filling $n=0.5$. The DSF from ED is nicely captured by the Gaussian approach, while the Bogoliubov dispersion deviates qualitatively. Since $t=J=0$, the ground state features tightly bound pairs with 
$v_k = \sqrt{n}$ and an analytical formula for the bright Gaussian dispersion, displayed by the black line, can be obtained (see the following subsection \ref{ssec:Simon}). Three flat bands of dark excited states also exist, the lowest one coming from  the $\beta(Q)$ part of the tangent space.
Finally, we remark that this set of
$U/P, V/P$ parameters 
is realizable within the Creutz ladder model introduced in Sec.~\ref{ssec:Creutz}; in other words, ($U/P, V/P$) lie on the white dashed line plotted in Fig.~\ref{fig:correls}.(a).

In panel (b), we consider a finite hopping $t/P=0.2$ that is still in the pair superfluid regime $(\beta=0)$. The dark flat bands found at $t=0$ acquire a finite bandwidth and (small) oscillator strength. In particular,  the highest band opens a gap in the main dispersion branch. The agreement between the ED spectrum and Gaussian state predictions is still excellent.

Finally, in panel (c), we investigate the superfluid regime ($\beta>0$) at $t/P=0.8$. In this case, the Bogoliubov and Gaussian approaches yield a close agreement with ED results at very low energies.
At larger energy, however, there are qualitative differences. We speculate that this is due to the sizable value of $U$, together with the 1D geometry which enhances strong correlations.

Overall, we conclude that the Gaussian ansatz is able to provide qualitatively accurate and  quantitatively good insights close to the pair condensate point, where the predictions of Bogoliubov theory can be completely wrong. When condensation occurs, both approaches provide similar estimates for the speed of sound.  
In Appendix \ref{app:correlated_hopping},
we discuss how superfluidity is driven
by the correlated hopping term $\hat{H}_J$.
At small densities,  it is found that $\hat{H}_J$ can drive condensation  only at large
values of $J$ and $U$, so that these condensates are highly correlated.
This suggests that, in flat bands, mean-field theory may capture the spectrum of $J$-driven superfluids  only at large densities.

Let us finally discuss the relation of our Gaussian state approach to the speed of sound that emerges in the (pair) superfluid state. In Refs.~\cite{Julku2021_excitations_of_BEC,Julku2021,Julku2023} it was shown that, under some assumptions, the Bogoliubov speed of sound is proportional to the square root of the quantum metric. While this connection is appealing, our Gaussian state analysis suggests that such simple result is not true in general. As detailed in Appendix \ref{app:dynamical_matrix}, the collective excitations of the Gaussian state are defined from the spectrum of a dynamical matrix $M(Q)$, whose matrix elements contain $h(k,p,q)$. So, it is more natural to think of the excitation modes as implicitly determined by the quantum geometric kernel. While this relation is in practice complicated and not very transparent, a simple limit is illustrated in the next paragraph. Here, we provide an explicit expression of the speed of sound as a function of the low harmonics $P,U,V$ of the quantum geometric kernel, holding for the PSF state in the absence of any single-particle hopping.

\subsection{Tight pair limit ($t=0$)}
\label{ssec:Simon}

As argued above, at small density the pair superfluid can be seen as a gas of bound pairs. As studied in Sec.~\ref{ssec:2body}, the competition between the pair hopping $P$ and the single-particle hopping $t$ determines the binding radius of the pairs. When $t=0$, the pairs are tightly bound, in the sense that the two bosons are forced to occupy the same site. In this case, the Gaussian ansatz for the ground state simplifies to the pair condensate wavefunction
\begin{equation}
    \ket{\Psi_g} = \hat{\mathcal{U}}_g |0\rangle =
    \exp \left( \frac{\lambda}{2} \sum_{j} 
    \left[ 
    (\hat{b}_{j}^\dagger)^2 -   \hat{b}_{j}^2 \right]
    \right)
    \ket{0},
\end{equation}
which has density $n=v^2$. The energy reads
\begin{equation}
    E/L = -\mu n +  (U-2P)u^2 v^2 + (2U + V) v^4,
\end{equation}
which leads to the saddle point equation
\begin{equation}
n = (\mu - U + 2P) / (6 U +2V - 4P),
\end{equation}
which provides the density. The pair coherence function is given by
\begin{equation}
     g^{(1)}_{\rm pair}(r) = 2\delta_{r,0} v^4 + (uv)^2 ,
\end{equation}
which scales like $n$ in the low density limit.

The  limit of tightly  bound pairs also allows one to obtain analytical information on the collective excitations, to which end we consider the Gaussian manifold 
\begin{align}
    \ket{\Psi} &= \hat{\mathcal{U}}_g \hat{\mathcal{U}} |0\rangle \nonumber \\
    &= \hat{\mathcal{U}}_g \exp \left( \frac{1}{\sqrt{2}} \sum_{j} 
    \left[ \lambda_j (\hat{b}_{j}^\dagger)^2 -  \lambda_j^* \hat{b}_{j}^2 \right]
    \right) \ket{0},
\end{align}
whose tangent space is spanned by the orthonormal basis 
\begin{equation}
    |j\rangle = \hat{\mathcal{U}}_g \hat{\mathcal{U}} \frac{(\hat{b}_{j}^\dagger)^2}{\sqrt{2}} \ket{0} \;.
\end{equation}
Applying the recipe of Appendix \ref{app:dynamical_matrix},
one finds one collective mode with dispersion
\begin{equation}
\omega(Q) = \sqrt{A^2(Q) - B^2(Q)}
\end{equation}
where
\begin{align}
    A(Q) &= 4 \Big[P + 3  (n + n^2) U \Big] \\
    & \qquad +  4 \Big[  (n + n^2) V - (1 + 2 n + 2 n^2) P\Big] \cos Q \\
    B(Q) &= 4 \left( n + n^2
    \right) \Big( 3U + V \cos Q - 2 P \cos Q \Big) .
\end{align}
This dispersion is plotted as the black line in Fig.~\ref{fig:spectra}.(a).
By expanding this at low momentum $Q \to 0$ we find an acoustic branch of dispersion
$\omega(Q) \simeq c |Q|$, with a speed of sound given by
\begin{equation}
    c_{\rm Gauss} = 4 \sqrt{n (1 + n) P (3U + V - 2 P)}.
\label{eq:cGauss}
\end{equation}
Notice that at the point $(U/P,V/P)=(0,2)$ one gets $c_{\rm Gauss}=0$, recovering the parabolic dispersion
expected for the pair condensate (see Appendix \ref{app:PairCondensate}).

The speed of sound 
obtained via the Gaussian ansatz in Eq. (\ref{eq:cGauss}) does not agree with 
the Bogoliubov-theory prediction (Appendix \ref{app:Bogo}). 
As a consequence, the result that the speed of sound is proportional to the square root of the quantum metric~\cite{Julku2021_excitations_of_BEC,Julku2021,Julku2023} does not hold  for the PSF,
while it is an excellent approximation in the SF regime.
Nonetheless, a partial analogy with those previous results can be established by noting that $c_{\rm Gauss}$  is determined by the low harmonics $P,U,V$ of the quantum geometric kernel [Eq.~\eqref{eq:QGK}].

\section{Conclusions and Outlook}
\label{sec:conclusions}

In summary,
we have studied the competition between pair superfluidity and standard superfluidity  by considering the effect of  single-particle processes in bosonic quasi-flat bands.
We have developed an effective description of pair superfluids and their excitation spectrum based on Gaussian states, benchmarking this approach through
analytical insights and exact diagonalization results. 
Crucially, the  Gaussian ground state wavefunction 
and 
the collective excitation spectrum
are expressed in terms of the quantum geometric kernel,
 which embodies the projection of the interactions on the quasi-flat band.
 In particular, the 
 speed of sound can be analytically extracted in the all-flat band limit and
 is not simply proportional to the quantum metric as predicted for single-particle condensates~\cite{Julku2021_excitations_of_BEC,Julku2021,Julku2023}.
Moreover, we have demonstrated in Appendix \ref{app:correlated_hopping} that correlated hopping can also drive the PSF to SF transition, in analogy to single-particle
hopping. However, $\hat{H}_J$ can also lead to collapse instabilities and 
strong correlations between next nearest-neighbor sites.

These methodological advances pave the way for future exciting investigations.
In this work, we limited ourselves to a one-dimensional system, in order to be able to use ED.
However, we expect  the Gaussian state ansatz to be even more
accurate in higher dimensions.
In this case, Quantum Monte Carlo and tensor networks may be used for the numerical benchmark.
Moreover, the Gaussian approach
is computationally cheap and can be applied to multi-orbital systems, without the need of performing any projection.
This can be important  when the dispersive bands touch the flat one, e.g. in kagome or Lieb lattices. 
The study of driven-dissipative extended Hubbard models represents another interesting direction.

On the experimental side, exciting goals for future research include the observation
of the PSF-SF phase transition, putatively driven by the Coleman-Weinberg mechanism~\cite{Coleman1973} or by fractional vortex nucleation at finite temperatures~\cite{Mukerjee2006},
and the direct
measurement of the collective excitation spectrum, with speed of sound tunable
through the Bloch geometry.
Fluxonium quantum simulators~\cite{Amelio2024Lasing}
and ultracold atomic systems~\cite{Sowiski2012} seem the most promising platforms.

\section*{Acknowledgements}

M.B. was financially supported by the fondation Roi Baudouin. I.A. and N.G. acknowledge financial support by the ERC grant LATIS, the EOS project CHEQS, the FRS-FNRS (Belgium)
and  the Fondation ULB.

~


\appendix

\section{Effect of correlated single-particle hopping $J$}
\label{app:correlated_hopping}

In the main text, we have discussed the competition between the PSF and SF phases
as a function of $t$, the single-particle hopping strength.
The goal of this Appendix is to show that correlated single-particle hopping contributions also promote the SF against the PSF.
This class of processes includes Hamiltonian terms such as 
$\hat{b}^\dagger_{j+1} \hat{n}_{j} \hat{b}_{j-1}, 
\hat{b}^\dagger_{j+1} \hat{b}_{j} \hat{n}_{j-1},
\hat{b}^\dagger_{j+1} \hat{b}^2_{j} \hat{b}^\dagger_{j-1},...$
and longer range contributions. These hoppings can arise in systems with isolated flat bands (e.g. the sawtooth chain).
In fact, these terms break the local symmetry of $\hat{H}_{PUV}$ and enable an extended single-particle coherence.
In spite of these analogies with $\hat{H}_t$, we will see that correlated hopping  can also drive collapse instabilities. Moreover, when considering
the specific case of $\hat{H}_J$ defined in Eq. (\ref{eq:HJ}),
coherences between next nearest neighbor sites
can be favored.
This is because 
$\hat{H}_{J} + \hat{H}_{PUV}$
separately conserves the parity of the total particle number on even and odd site sublattices. 
These correlations
are qualitatively captured by the Gaussian state approach; in fact, we have picked up
this example of correlated hopping term so to show the power of the method.

\subsection*{ED phase diagram}

\begin{figure*}[t]
\includegraphics[width=0.9\linewidth]{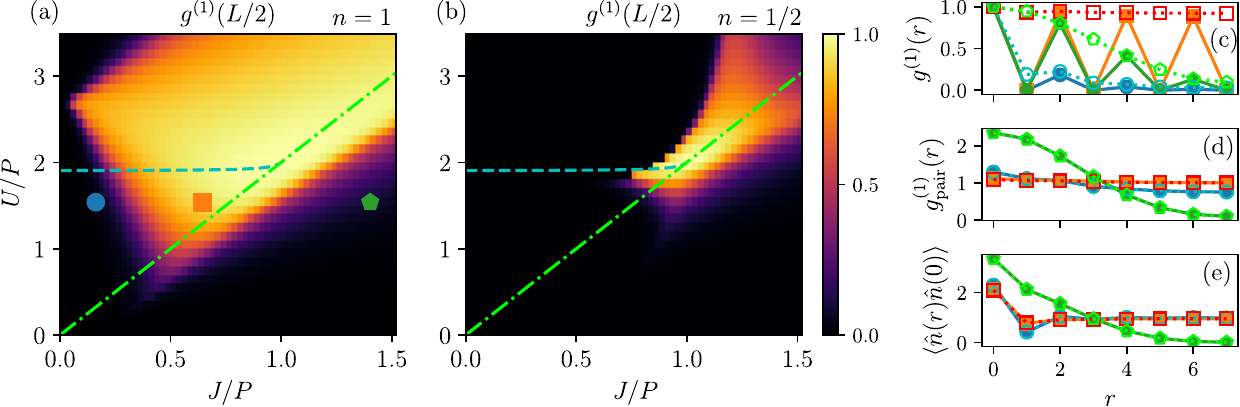}
    \caption{
    The effect of the correlated hopping $J$ and the onsite interaction strength $U$ on the coherence of the system is investigated, for fixed $V=2P$.
    In panel (a), we 
    display $g^{(1)}(L/2)$ at
    unit filling $L=N=14$, while half filling 
     $L=16,N=8$ is shown in panel (b).
    The cyan dashed line corresponds to
    the binding threshold of Eq. (\ref{eq:2body_tUV_line}),
    defined via the 2-body problem.
    The collapse instability condition
    $h(0,0,0)=0$
    is indicated by the green dashed-dotted line.
    In panel (c,d,e), we respectively report the single-particle coherence function, the pair coherences and density-density correlator as a function of distance $r$,
    for the three different points indicated in panel (b).
    The PSF is indicated by the blue circle,
 the SF  by the orange square,
and the clustered phase by the green pentagon.
The empty versus filled symbols correspond to $t=0.05P$ and $t=0$, respectively.
    }
    \label{fig:J_ED}
\end{figure*}

In Fig.~\ref{fig:J_ED}, we assess via ED the impact of the correlated hopping term $J$. In panel (a), we consider the single-particle coherence by plotting $\gsp(L/2)$ as a function of $J/P$ and $U/J$, for fixed $V=2P$, negligible single-particle hopping $t=0.05P$ and unit filling $N=L=14$.
One can observe some analogy with Fig.~\ref{fig:t_ED}: in particular,  increasing $J$ favors the SF phase over the PSF. However, the correlated single-particle hopping term is a two-particle operator, and, while favoring hopping, it can lead to collapse instabilities if it is not counteracted by the $U,V$ contributions.
The collapse instability results in a rapid decay of the long-range coherence. 
In practice, we find that the instability qualitatively occurs when $h(0,0,0)=U+V-2P-2J$ becomes negative, corresponding to the mean-field potential becoming attractive. The $h(0,0,0)<0$ region corresponds to the area below the green dashed-dotted line. The dashed cyan line corresponds to the binding threshold of Eq. (\ref{eq:2body_tUV_line}), from the 2-body calculation of Sec.~\ref{ssec:2body} below.
At unit filling, this line does not match quantitatively with the PSF to SF transition.

In Fig.~\ref{fig:J_ED}.(b), we instead consider half filling with $L=16,N=8$.
At this lower density, the two-body results are more compelling and we expect PSF below the dashed cyan line. Moreover, the stability condition $h(0,0,0)>0$ must also be required (region above green dashed-dotted line).
This leads to a sizable reduction of the region where the SF phase dominates.
In particular, 
since a large $U/P$ is needed in order to stabilize the SF,
one can conclude that at small densities there is no weakly interacting SF phase purely driven by correlated hopping. As a consequence, the validity of single-particle mean-field theory for flat-band superfluids needs to be critically assessed in future studies.

In Fig.~\ref{fig:J_ED}.(c-e), we report
$g^{(1)}(r), \ g^{(1)}_{\rm pair}(r)$ and $\langle \hat{n}(0) \hat{n}(r) \rangle$
in the unit filling case for the three points indicated in panel (a).
The empty symbols correspond to a small but finite $t=0.05$, while full symbols to stricly zero $t=0$.
 In fact, the $J$ term induces hopping between next-nearest neighbor sites, so that, for $t=0$,  
$\hat{H}_J + \hat{H}_{PUV}$
conserves the total particle number of even and odd sites separately.
The ground state can then be seen as a Schr\"odinger cat superposition of two quasi-condensate states, with aligned or anti-aligned phases between nearest-neighbor sites. 
However, even a small $t$ is enough to synchronize the two quasi-condensates, suppressing the anti-aligned component.
This behavior is clearly visible in panel (c),
which displays a staggered pattern for $t=0$, while a small $t$ has a negligible effect in panels (d) and (e).

The blue circle corresponds to a PSF, with exponentially decaying $g^{(1)}(r)$
and algebraic $g^{(1)}_{\rm pair}(r)$.
The SF is instead given by the orange squares, while the green pentagon indicates the collapsed phase, where the bosons cluster and therefore the density-density correlator decays at large separations.

\subsection*{Gaussian ground state }

\begin{figure}[t]
    \centering
\includegraphics[width=0.9\linewidth]{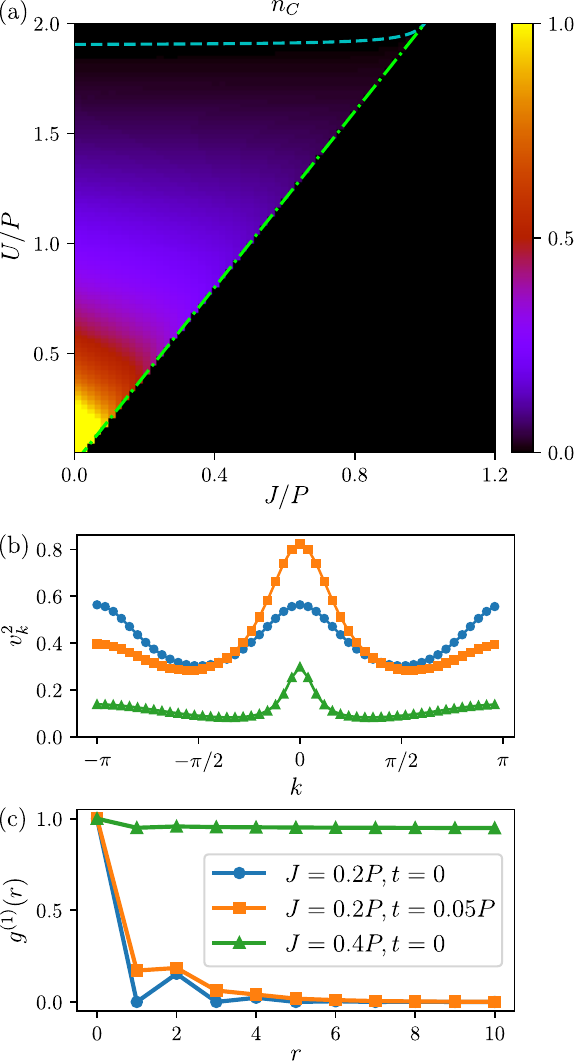}
    \caption{ Gaussian ground state and correlated hopping.
    (a) The critical density as a function of $J/P$ and $U/P$, for fixed $V=2P, t=0.05P$. The green dashed-dotted line denotes the onset of collapse instability, while the cyan dashed line refers to the 2-body binding threshold of Eq. (\ref{eq:2body_tUV_line}).
    (b) Variational parameters $v_k$
    and (c) coherence function 
    $g^{(1)}(r)$ for three specific choices of $t,J$, at fixed $\mu=-0.3P$.
    }
    \label{fig:Gaussian_J}
\end{figure}

Next, we have studied the behavior of the ground state with $J$ within the Gaussian state approach. This subsection parallels Sec.~\ref{ssec:GaussGS} , where single-particle hopping was considered.
The maximal PSF density $n_C$ 
is plotted in Fig.~\ref{fig:Gaussian_J}.(a) as a function of $U/P$ and $J/P$, for a small $t=0.05$.
The 2-body binding threshold is provided by the cyan dashed line, which is exactly recovered in the low density regime by the Gaussian state ansatz.
Correlated hopping can also give rise to the collapse instability for $h(0,0,0)<0$, as delimited by the green dashed dotted line.

The correlation functions reproduce qualitatively the ED results.
In particular, 
as visible in panel (c),
the PSF (blue line) displays a staggered $g^{(1)}(r)$ for $t=0$, which is smoothened by introducing a small $t$ (orange). 
The staggering pattern derives from the double peak structure of 
 $v_k$,  depicted in panel (b).

\subsection*{Dynamic structure factor}

\begin{figure}[t]
    \centering
\includegraphics[width=0.99\linewidth]{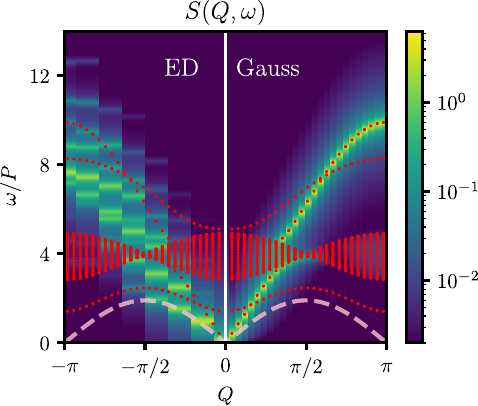}
    \caption{ 
    The dynamic structure factor $S(Q,\omega)$ was computed in both exact diagonalization (displayed for $Q \leq 0$) and the Gaussian state approach (in the $Q \geq 0$ half), for the pair superfluid in the presence of $\hat{H}_J$.
    The red dots denote the eigenvalues of the Gaussian dynamical matrix $M(Q)$,
    while the pink dashed line is the Bogoliubov dispersion from Appendix \ref{app:Bogo}.
    As parameters we used $J=0.08P,t=0, U=0.1P, V=2.1P,n=1.0$
    }
    \label{fig:spectrumJ}
\end{figure}

Finally, in Fig.~\ref{fig:spectrumJ},
we study the 
dynamic structure factor of the
PSF with $J=0.08P, U=0.1P,V=2.1P,t=0$.
The shape of the ED dispersion is roughly captured by the Gaussian approach.
It seems that, if  the lowest and highest weakly dispersive
bands opened some gap in the main (bright) dispersion branch,  the Gaussian spectrum would resemble even more closely the ED findings.
The Bogoliubov prediction is instead rather inaccurate.

\section{Pair condensation: an exact result}
\label{app:PairCondensate}

As proven below, it turns out that the {\em pair condensate} state
\begin{equation}
    |\Psi\rangle_{\text{PC}}
    =
    \frac{1}{Z_N}
    \left(
\hat{\Pi}_0^\dagger
    \right)^{N/2}
    |0\rangle,
\end{equation}
with the pair creation operator of total momentum $Q$
\begin{equation}
    \hat{\Pi}_Q^\dagger =  \sum_j e^{iQj}(\hat{b}_j^\dagger)^2
\end{equation}
and $Z_N$ a normalization factor, is the exact ground state of $\hat{H}_{PUV}$ for $V=2P>0$ and $U=0$, in the $N$ particle sector (with $N$ even). This state $|\Psi\rangle_{\text{PC}}$ contains $N/2$ zero momentum pairs, and is the pair-condensate equivalent of a pure single-particle condensate state. Notice that, in complete analogy to non-interacting single-particle condensates, the system is not superfluid for this precise parameter set, as it turns out that the excitation spectrum is gapless and parabolic. This exact solution is accessed within the Creutz ladder model for attractive intra-rung interaction $U_1=-U_0<0$.

The proof of the above statements is based on rewriting the Hamiltonian for $V=2P$ and $U=0$ as
\begin{equation}
    \hat{H}_{PUV}(V=2P,U=0) = P \sum_j \hat{B}_{j,j+1}^2 - 2P\hat{N},
\end{equation}
where we introduced the local current operator
\begin{equation}
    \hat{B}_{j,j+1} = i \left( \hat{b}^\dagger_j \hat{b}_{j+1} -
\hat{b}^\dagger_{j+1} \hat{b}_j
\right),
\end{equation}
which is a hermitian operator. This  operator commutes with the pair creation operator
\begin{equation}
    \left[ \hat{B}_{j,j+1}, \hat{\Pi}_0^\dagger\right] = 0, \qquad \forall j
\end{equation}
entailing indeed that 
\begin{equation}
\sum_j \hat{B}_{j,j+1}^2  |\Psi\rangle_{\text{PC}} = 0 \;.
\end{equation}
Since $P \sum_j \hat{B}_{j,j+1}^2$ is a non-negative operator, 
$|\Psi\rangle_{\text{PC}}$ is the ground state in the $N$ particle sector, with energy $-2NP$. Moreover,  a single pair of momentum $Q$ is an exact eigenstate of $\hat{H}_{PUV}(V=2P,U=0)$ with energy $-4P\cos Q$ (see also Sec. \ref{ssec:2body} for the mapping of the 2-body problem to a single-particle problem). As a consequence,  the state 
\begin{equation}
    \ket{Q} = \left( \hat{\Pi}_0^\dagger \right)^{N/2-1} \hat{\Pi}_Q^\dagger |0\rangle
\end{equation}
is an eigenstate of momentum $Q$ and energy 
\begin{equation} 
\omega(Q) =  -2PN + 4P(1- \cos Q) \simeq  2P (-N + Q^2)
\end{equation}
the last approximation holding for small $Q$. In other words, the neutral excitation spectrum is gapless and parabolic in the momentum, so that the Landau criterion rules out superfluidity. As a final remark, notice
that all these arguments hold for any dimensionality of the lattice.

\section{Single-particle mean-field and Bogoliubov modes}
\label{app:Bogo}

Previous studies have investigated Bose-Einstein condensation in flat bands  relying on single-particle mean-field theory~\cite{Julku2021_excitations_of_BEC, Julku2021, Julku2023, Amelio2024Lasing}.
This can provide reasonable results if correlated hopping favors the superfluid phase over pair superfluidity.
It was demonstrated that a non-zero quantum metric~\cite{Yu2025} can stabilize the condensate and endow the Bogoliubov excitation spectrum with a finite speed of sound.
In this Appendix, we provide a brief reminder of the Bogoliubov results.

In the single-particle mean-field theory~\cite{Pitaevskii2016}
the bosonic operators are expanded as
\begin{equation}
\hat{b}_Q = \delta_{Q,0} \sqrt{n}  + \delta \hat{b}_Q.
\end{equation}
Keeping up to quadratic terms in the Hamiltonian, one can derive the Heisenberg equation
\begin{equation}
    i \partial_t \delta \hat{b}_Q
    =
    A(Q) \delta \hat{b}_Q
    + B(Q)  \delta \hat{b}^\dagger_{-Q},
\end{equation}
with 
\begin{align}
A(Q) &= \epsilon_Q + 2 [h(Q,0)+h(Q,Q)] n     \\
B(Q)&=2 h(0,Q) n\;.
\end{align}
The chemical potential is fixed by the condition
$\epsilon_0 + 2 h(0,0) n = 0 $.
The Bogoliubov mode frequency then reads:
\begin{equation}
    \omega_{\rm Bogo}(Q)
    =
     \sqrt{ A^2(Q)
    -
    B^2(Q) }.
\end{equation}
The small momentum expansion
leads to the speed of sound
\begin{equation}
    c_{\rm Bogo}
    =
    \sqrt{4n (U+V-2P-2J) [
t + 4n(P+J)
    ]}.
    \label{eq:cBogo}
\end{equation}

Previous works have demonstrated that, in flat band systems and under some assumptions, $c_{\rm Bogo}$
is proportional to the square root of the quantum metric~\cite{Julku2021,Julku2021_excitations_of_BEC,Julku2023,Amelio2024Lasing}.
Indeed, for  on-site Hubbard interactions $U_{\alpha \beta }=U_0 \delta_{\alpha\beta}$
 ~\footnote{This is not the case in the Creutz ladder proposal of Sec.~\ref{ssec:Creutz}, since $U_0 = - U_1$.
Note the same $h$ may correspond to different microscopic models and quantum metrics.}, 
the  ``condensate quantum distance''  $d(k) = \sqrt{1 - h(k,0,0)/h(0,0,0)}$, 
was introduced~\cite{Julku2021_excitations_of_BEC,Julku2023}.
The corresponding   metric $g$ is  defined through
$d^2(k) = g k^2 + O(k^3)$,
so that, in the flat-band limit $t=0$, Eq. (\ref{eq:cBogo}) can then be recast into
$c_{\rm Bogo} = 4 h(0,0) n \sqrt{|g|}$,
with $g=-(P+J)/(U+V-2P-2J)$.

In this work, we used the Gaussian state variational approach to derive more accurate predictions for the collective excitation spectrum,
holding in both the SF and PSF regimes.
While it is clear that the geometry of the Bloch functions plays a central role,
we found more natural and general to relate the collective modes and speed of sound 
$c_{\rm Gauss}$
to the quantum geometric kernel (or its low harmonic expansions), rather than to the quantum metric.
As analyzed in Fig.~\ref{fig:spectra},
in the pair superfluidity regime
Bogoliubov theory leads to markedly different predictions than the Gaussian state approach and ED,
while it matches at low energies with the Gaussian state approach when single-particle condensation sets in, i.e. for $\beta \neq 0$.
At the pair condensate point $(V/P,U/P)=(2,0)$ with $t=0$ and $J=0$, we find $\omega_{\rm Bogo}(Q)=0$ instead of the parabolic spectrum expected from  ED and recovered by the Gaussian approach (not shown). 
Moreover, the functional dependence on the density $n$ is different within the two methods.
In the dispersive Bose-Hubbard model scenario ($J=P=V=0$),
we recover the usual behavior
$c_{\rm Bogo} = 2 \sqrt{n U t}$~\cite{Guaita2019}.
In the flat-band case, the dependence on the Hartree-shift $h(0,0) n$ is instead linear.
In contrast, the  Gaussian state approach predicts $c_{\rm Gauss} \propto \sqrt{n(1+n)}$.

\section{Dynamical matrix and linear response}
\label{app:dynamical_matrix}

In this Appendix, we provide more details on the derivation of the energy $E_g$, as well as of the dynamical  matrix $M(Q)$ given in Eq. \eqref{eq:M_matrix} and the density-density linear response $\chi(Q,\omega)$ of Eq. \eqref{eq:GaussDSF}. 

First, it is useful to expand $\hat{H}_g \equiv\hat{\mathcal{U}}_g^\dagger \hat{H} 
\hat{\mathcal{U}}_g$ through
the rotation of Eq. (\ref{eq:Bogo_rotation}). 
One can then explicitly calculate the expectation
the energy functional
\begin{equation}
    E_g(\vec{x}_g) =
\langle \Psi_g(\vec{x}_g) |
\hat{H}
|\Psi_g(\vec{x}_g)\rangle = \bra0 \hat{H}_g \ket0
\end{equation}
by means of Wick's theorem, leading to Eq. \eqref{eq:GS_energy}.
By minimizing the energy, one finds the ground state parameters $\beta$ and $\lambda_k$.

On top of this Gaussian ground state, one can derive the collective excitations and the linear response to a density perturbation.
Specifically,
we define the perturbed, time-dependent Hamiltonian 
\begin{equation}
    \hat{H}_\epsilon(t)
=
\hat{H} + \epsilon
e^{\eta t} \big( \hat{W} e^{-i \omega t} + \hat{W}^\dagger e^{i \omega t} \big)
\end{equation}
where $\epsilon$ is the strength of the perturbation, $\eta$ regularizes the response by turning off the perturbation at $t=-\infty$, and  
$\omega$ describes the frequency of the drive. 
The drive operator 
$\hat{W}$ will be eventually chosen to be the density modulation at momentum $Q$,
$\hat{W} = \hat{n}(Q) - n \delta_{Q,0}  = \frac{1}{L}\sum_k \hat{b}^\dagger_{k+Q} \hat{b}_k
- n \delta_{Q,0}$.

We parametrize  the
Gaussian state manifold as
\begin{equation}
    |\Psi(\vec{x})\rangle
    =
    \hat{\mathcal{U}}_g(\vec{x}_g)
    \hat{\mathcal{U}}(\vec{x})
    | 0 \rangle, 
    \label{eq:parametrization}
\end{equation}
with $\hat{\mathcal{U}}(\vec{x})
=
\hat{\mathcal{D}}(\{ \beta(Q) \})
\hat{\mathcal{S}}( \{ \lambda_k(Q) \})
$
and
\begin{equation}
    \hat{\mathcal{D}}(\{ \beta(Q) \}) = \exp \left( 
    \sum_Q \beta(Q) \hat{b}_Q^\dagger - \beta^*(Q) \hat{b}_Q \right),
\end{equation}
\begin{equation}
    \hat{\mathcal{S}}( \{ \lambda_k(Q) \}) = \exp \left( 
    \sum_{kQ}
    \frac{
    \lambda_k(Q) \hat{b}_k^\dagger
    \hat{b}_{Q-k}^\dagger
    - 
    \lambda^*_k(Q) \hat{b}_k
    \hat{b}_{Q-k}
    }
    {
\sqrt{1 + \delta_{k, Q-k}}
    }
    \right).
\end{equation}
The sum over $k$ is restricted so to avoid the overparametrization arising from 
$\hat{b}_k^\dagger
    \hat{b}_{Q-k}^\dagger
    =\hat{b}_{Q-k}^\dagger \hat{b}_k^\dagger
    $; in practice, we choose the constraint $k \geq Q-k$, where momenta are meant to be within the first Brillouin zone.
    At this level, we keep $\hat{W}$ generic and sum over all momenta $Q$.
The square root is a normalization factor that will ensure the orthonormality of the tangent vector basis introduced below.

In the following, we take inspiration from the geometric treatment of Gaussian states presented in Refs.~\cite{Guaita2019,Hackl2020}. 
The Dirac time-dependent variational principle~\cite{Dirac_1930}
leads to the projected Schr\"odinger equation
\begin{equation}
    i \hat{\mathbb{Q}}  \partial_t \ket{\psi}
    =
    \hat{\mathbb{P}} \hat{H}_\epsilon(t) \ket{\Psi},
\end{equation}
where $\hat{\mathbb{Q}}$ implements the projection to the linear space normal to 
$\ket{\Psi}$, and it is necessary because the variational ansatz does not allow for phase or normalization factors, and thus corresponds to a projective representation.
The operator $\hat{\mathbb{P}}$, instead, projects on the tangent space to the variational manifold in $\ket{\Psi}$.
Using the chain rule leads to
\begin{equation}
   i  \sum_{a} \dot{x}^a  \ket{a}
    =
    \hat{\mathbb{P}}  \hat{H} \ket{\Psi},
    \label{eq:ixmu}
\end{equation}
where 
$a$ labels the variational parameters $\vec{x} = \{ \beta(Q), \lambda_k(Q) \}$
and the tangent space vectors
$\ket{a} = \hat{\mathbb{Q}} \  \partial_a  \ket{\psi}$
were introduced.
Notice that the tangent space is
a complex vector space, since the manifold of Gaussian states possesses a K\"ahler geometric structure. In particular, the imaginary unit commutes with the projector $\mathbb{P}$.
In the non-K\"ahler case,
one should treat separately the real and imaginary parts of the variational parameters and view the tangent space as a real vector space.
In order to invert this equation, one would introduce
the manifold metric
$g_{ab}=2{\rm Im}\langle a | b \rangle$ and its inverse $G^{ab}$, so to obtain $ \dot{x}^a  
    =
   2 \sum_b G^{ab}  {\rm Re}\bra{b} \hat{H} \ket{\psi}$.

The parametrization in Eq. (\ref{eq:parametrization})
provides two advantages.
First, separating $\vec{x}_g$ and $\vec{x}$ guarantees that  
the $\vec{x}$ are of order $\epsilon$, with $\vec{x}=\vec{0}$
corresponding to the ground state 
$|\Psi_g \rangle$.
Second, by factorizing 
$\hat{\mathcal{U}}_g$ and $\hat{\mathcal{U}}$
we can avoid computing some commutators.
Leaving out corrections of order $\epsilon^2$, we thus find that the tangent space is spanned by the orthonormal basis of vectors
\begin{equation}
    \ket{Q} = \hat{\mathcal{U}}_g \hat{\mathcal{U}} \hat{b}^\dagger_Q \ket{0} + O(\epsilon^2),
\end{equation}
\begin{equation}
    \ket{k,Q} = \hat{\mathcal{U}}_g \hat{\mathcal{U}} \frac{\hat{b}^\dagger_k \hat{b}^\dagger_{Q-k}}{\sqrt{1+\delta_{k,Q-k}}}
    \ket{0}   + O(\epsilon^2),  \ \ \ {\rm with} \ k \geq Q-k.
\end{equation}
Relying on this orthonormality,
we can simply invert Eq. (\ref{eq:ixmu}) to obtain
\begin{equation}
   i  \dot{x}^a  
    =
    \bra{a} \hat{H}_\epsilon(t) \ket{\Psi} + O(\epsilon^2).
    \label{eq:ixmu_inv}
\end{equation}
Notice that the saddle point equations are equivalent to
the zeroth order of this equation
$\bra{a} \hat{H} \ket{\Psi}=0$.
At order $\epsilon$
and focusing on the drive $\hat{W}$ at a specific momentum $Q$,
we thus find a linear set of equations of motion, coupling 
$\{ \beta(Q), \lambda_k(Q) \}$
and 
$\{ \beta^*(-Q), \lambda^*_k(-Q) \}$.
To deal with the particular structure of these equations,
it is convenient to go to frequency space and write them as
\begin{equation}
    (\omega + i\eta)
    \vec{X}(Q,\omega) 
    = M(Q)     \vec{X}(Q,\omega) 
    +
    \epsilon \sigma_z \vec{F},
    \label{eq:EOM_X}
\end{equation}
where  
$\vec{X}(Q,\omega)$
is the  Fourier component at frequency $\omega + i\eta$ of
$\vec{X}(Q,t)  = \left(
    \beta(Q),
        \vec{\lambda}(Q),
        \beta^*(-Q),
        \vec{\lambda}^*(-Q)
    \right)^T$. The first (second) half of 
    $\vec{X}$ is commonly called the particle (hole) component.
    The matrix $\sigma_z$ acts as the identity on the particle component, and as minus the identity on the hole half. 
    We also introduced the particle-hole symmetric   matrix
\begin{equation}
M(Q)
=
     \begin{pmatrix}
        A & B \\
        - B^\dagger & -A^\dagger
    \end{pmatrix}.
    \label{eq:M_matrix}
\end{equation}
In our case, $A=A^\dagger$ is hermitian.
The source term is  given by
$\vec{F} = (\vec{f}^{(-)}, \vec{f}^{(+)})^T$,
with
matrix elements $f^{(-)}_a = \bra{a} 
     \hat{W}
     \ket{\Psi_g}|_{\epsilon=0}$
     and 
     $f^{(+)}_a = \bra{a} 
     \hat{W}^\dagger
     \ket{\Psi_g}|_{\epsilon=0}$,
     computed at zero order in $\epsilon$ (i.e. setting $\hat{\mathcal{U}}=1$ in $|a\rangle$).
The matrix elements of $A,B,F$
can  be calculated 
combining 
the rotation of Eq. (\ref{eq:Bogo_rotation})
with Wick theorem.
The detailed derivation and the resulting expressions are reported at the end of this Appendix.

As a consequence of the particle-hole symmetry,  the eigenvalues   of $M(Q)$ can be ordered into the vector $\vec{\Omega}=(\vec{\omega},-\vec{\omega})^T$, with all the elements of $\vec{\omega}$ chosen to be non-negative~\footnote{This is is true for $Q \neq 0$ and in the absence of dynamical instabilities (associated with complex frequencies, occurring in conjugate pairs). For $Q=0$, the Goldstone mode has zero frequency and is not normalizable}.
Furthermore, within this ordering,
the modes can be normalized so that
$L^\dagger \sigma_z L = \sigma_z$,
where the matrix $L$ collects the eigenmodes of $M(Q)$, according to $M=L \ {\rm diag}(\vec{\Omega}) L^{-1}$.
Inverting Eq. (\ref{eq:EOM_X}), we thus find
\begin{align}
    \vec{X}(Q,\omega) 
    &=   \epsilon\frac{1}{\omega - M + i\eta}
    \sigma_z \vec{F} \nonumber \\
    &=    \epsilon L \frac{ \sigma_z}{\omega - {\rm diag}(\vec{\Omega}) + i\eta}
     L^\dagger \vec{F},
\end{align}
where in the last equality we used that $L^\dagger \sigma_z L = \sigma_z$ entails
$ L^{-1} \sigma_z = \sigma_z L^\dagger$.

The density-density response function $\chi$ of the system is then defined by considering the expectation value~\cite{Pitaevskii2016}
\begin{equation}
   \langle \Psi |
   \hat{W}^\dagger |
   \Psi
   \rangle 
   =
     \vec{F}^\dagger 
    \cdot
    \vec{X}(Q) + O(\epsilon^2),
    \label{eq:W_expct_value}
\end{equation}
whose $\epsilon  e^{-i(\omega+i\eta)t}$ component  is set equal to $\chi(Q,\omega)$.
Altogether, it is found
\begin{equation}
    \chi(Q,\omega)
    =
    \sum_s
    \frac{{\rm sign}(\Omega_s)}{\omega - \Omega_s + i\eta}
    |(L^\dagger \vec{F})_s|^2. 
    \label{eq:GaussDSF}
\end{equation}
Finally, the dynamic structure factor is given by
$
S(Q,\omega) =
-2{\rm Im} \ \chi(Q,\omega)$.

\onecolumngrid

\subsection*{Explicit equations of motions and matrix elements of $M(Q)$}

At order $\epsilon$, Eq. (\ref{eq:ixmu_inv})
can be explicitly written as the following set of equations:
\begin{equation}
\begin{split}
        i \dot{\beta}(Q)
     & =
   A_{\beta\beta}(Q) \beta(Q) 
     + 
    \sum_{k \geq Q-k} 
   A_{\beta\lambda_k}(Q)
   \lambda_k(Q)  
    +
   \sum_{-k \geq -Q+k}
   B_{\beta\lambda^*_{-k}}(Q)
   \lambda^*_{-k}(-Q)
    + \,
   \epsilon e^{\eta t} [ e^{-i\omega t} f^-_{\beta}(Q)
   +e^{i\omega t} f^+_{\beta}(Q)],
   \\
i \dot{\lambda}_k(Q)
    & =
    A_{\lambda_k\beta}(Q)
   \beta(Q)  +  
   B_{\lambda_k\beta^*}(Q)
   \beta^*(-Q)
    +
   \sum_{p \geq Q-p} A_{\lambda_k\lambda_p}(Q) \lambda_p(Q) \\
    & +
    \sum_{-p \geq -Q+p} B_{\lambda_k\lambda^*_{-p}}(Q) \lambda_{-p}(-Q)
     + \,
   \epsilon e^{\eta t} [ e^{-i\omega t} f^-_{\lambda_k}(Q)
   +e^{i\omega t} f^+_{\lambda_k}(Q)],
    \end{split}
\end{equation}
with 
\begin{equation}
    A_{\beta\beta}(Q) = \left[ 
    \bra{0} \hat{b}_Q \hat{H}_g \hat{b}^\dagger_Q \ket{0}
    - E_g
    \right]
\end{equation}
\begin{equation}
    A_{\beta\lambda_k}(Q)
=
\frac{
\bra{0} \hat{b}_Q  \hat{H}_g \hat{b}^\dagger_{k} \hat{b}^\dagger_{Q-k} \ket{0}
   }{
   \sqrt{1+\delta_{k,Q-k}}
   }
\end{equation}
\begin{equation}
    A_{\lambda_k\lambda_p}(Q)
   =
   \frac{
[\bra{0} \hat{b}_{k} \hat{b}_{Q-k}   
\hat{H}_g 
\hat{b}^\dagger_{p} \hat{b}^\dagger_{Q-p} \ket{0} - E_g (\delta_{k,p} + \delta_{Q-k,p})]
   }{
   \sqrt{1+\delta_{k,Q-k}}
   \sqrt{1+\delta_{p,Q-p}}
   } 
\end{equation}
\begin{equation}
    B_{\beta\beta^*}(Q) = 
    \bra{0} \hat{b}_Q \hat{b}_{-Q} \hat{H}_g  \ket{0}
\end{equation}
\begin{equation}
    B_{\lambda_{k}\beta^*}(Q)
    =
   \frac{
\bra{0} \hat{b}_{k} \hat{b}_{Q-k} \hat{b}_{-Q}  \hat{H}_g  \ket{0}
   }{
   \sqrt{1+\delta_{k,Q-k}}
   }
\end{equation}
\begin{equation}
    B_{\lambda_{k}\lambda_{-p}^*}(Q)
   =
   \frac{
\bra{0} \hat{b}_{k} \hat{b}_{Q-k}   
\hat{b}_{-p} \hat{b}_{-Q+p}
\hat{H}_g 
 \ket{0}
   }{
   \sqrt{1+\delta_{k,Q-k}}
   \sqrt{1+\delta_{-p,-Q+p}}
   }.
\end{equation}
A compact representation of these equations gives the particle-hole symmetric dynamical matrix
\begin{equation}
M(Q) = 
    \begin{pmatrix}
        A & B \\
        -B^\dagger & -A^\dagger
    \end{pmatrix},
\end{equation}
where $A = \begin{pmatrix}
    A_{\beta \beta}(Q) & A_{\beta \lambda_k}(Q) \\
    A_{\lambda_k \beta}(Q) & A_{\lambda_k \lambda_p}(Q)
\end{pmatrix}$ and $B = \begin{pmatrix}
    B_{\beta \beta^*}(Q) & B_{\beta \lambda_{-k}^*}(Q) \\
    B_{\lambda_k \beta^*}(Q) & B_{\lambda_k \lambda_{-p}^*}(Q)
\end{pmatrix}$.
Notice that because of inversion symmetry the off-diagonal blocks are simply the transpose of one another.
To ensure the explicit particle-hole structure of the dynamical matrix $M(Q)$,
the ordering convention of the indices must be consistent between the momenta of the particle and hole sectors. 
A possible choice is that the index $k$ of $\lambda_k(Q)$ come in increasing order (over the BZ), while the $-p$ index of $\lambda^*_{-p}(-Q)$
be decreasing (in such a way, the first contributions correspond to $k=k_1$ and $-p=-k_1$, respectively). 
Using Wick's theorem, we explicitly calculate the particle-particle terms in $A(Q)$ as
\begin{equation}
     \bra{0} \hat{b}_Q \hat{H}_g \hat{b}^\dagger_Q \ket{0} - E_g = \frac{\mathcal{E}_Q}{u_Q^2 + v_Q^2} 
\end{equation}
with 
$\mathcal{E}_Q = 
\varepsilon_Q + 2\beta^2\Big(h(Q,Q) + h(Q,0) \Big) + \Xi(Q)$, 
together with
\begin{equation}
    \begin{split}
        \bra{0} \hat{b}_Q  \hat{H}_g \hat{b}^\dagger_{k} \hat{b}^\dagger_{Q-k} \ket{0}/2\beta 
        & = 
         u_{Q}u_{Q-k}v_{k}\Big[ h(Q-k,k) + h(Q-k,Q) \Big] 
          + u_{k}v_{Q}v_{Q-k} \Big[ h(Q-k,k) + h(Q-k,Q) \Big] 
         \\ 
         &
         + u_{Q}u_{k}v_{Q-k}\Big[ h(k,Q) + h(k,Q-k) \Big] 
          + u_{Q-k}v_{Q}v_{k}
          \Big[ h(k,Q) + h(k,Q-k) \Big] 
         \\
         &
         + u_{Q}u_{Q-k}u_{k}\Big[ h(Q,k) + h(Q,Q-k) \Big]  
         + v_{Q}v_{Q-k}v_{k} 
         \Big[ h(Q,k) + h(Q,Q-k) \Big], 
    \end{split}
\end{equation}
\begin{equation}
    \begin{split}
 & \ \ \ \ \bra{0} \hat{b}_{k} \hat{b}_{Q-k} \hat{H}_g \hat{b}^\dagger_{p} \hat{b}^\dagger_{Q-p} \ket{0} - E_g (\delta_{k,p} + \delta_{Q-k,p}) =  \big(\delta_{k,p} + \delta_{Q-k,p} \big) \big( 
    \frac{\mathcal{E}_k}{u_k^2 + v_k^2} + \frac{\mathcal{E}_{Q-k}}{u_{Q-k}^2 + v_{Q-k}^2}  \big)  
    \\
     + & 2\bigg\{ u_{Q-k}u_{k}u_{Q-p}u_{p} \Big[ h(Q, p-k) + h(Q, Q-p-k) \Big] 
     +  v_{Q-k}v_{k}v_{Q-p}v_{p} \Big[ h(Q, p-k) + h(Q, Q-p-k) \Big] 
     \\
     + &   u_{p} u_{k} v_{Q-k} v_{Q-p} \Big[ h(Q-p-k, Q) + h(Q-p-k, p-k) \Big] 
     + u_{Q-p} u_{Q-k} v_{k} v_{p} \Big[ h(Q-p-k, Q) + h(Q-p-k, p-k) \Big] 
     \\
     +  & u_{Q-p} u_{k} v_{Q-k} v_{p} \Big[ h(p-k, Q) + h(p-k, Q-p-k) \Big] 
     +  u_{p} u_{Q-k} v_{k} v_{Q-p} \Big[ h(p-k, Q) + h(p-k, Q-p-k) \Big] 
      \bigg\}
    \end{split}
\end{equation}
The particle-hole block $B(Q)$ instead reads as follows:
\begin{equation}
    \bra{0} \hat{b}_Q \hat{b}_{-Q} \hat{H}_g  \ket{0}
     =
     0
\end{equation}
\begin{equation}
    \begin{split}
        \bra{0} \hat{b}_Q \hat{b}_{-k} \hat{b}_{-Q+k} \hat{H}_g  \ket{0}/2\beta  & = 
        u_{Q}u_{k}v_{Q-k} \Big[ h(Q-k,k) + h(Q-k,Q) \Big] 
        + u_{Q-k}v_{Q}v_{k} \Big[ h(Q-k,k) + h(Q-k,Q) \Big] 
        \\ &
        + u_{Q}u_{Q-k}v_{k} \Big[ h(k,Q) + h(k,Q-k) \Big] 
        + u_{k}v_{Q}v_{Q-k} \Big[ h(k,Q) + h(k,Q-k) \Big] 
        \\ &
        + u_{Q-k}u_{k}v_{Q} \Big[ h(Q,k) + h(Q,Q-k) \Big] 
        + u_{Q}v_{Q-k}v_{k} \Big[ h(Q,k) + h(Q,Q-k) \Big]
    \end{split}
\end{equation}
\begin{equation}
    \begin{split}
& \bra{0} \hat{b}_{k} \hat{b}_{Q-k}  \hat{b}_{-p} \hat{b}_{-Q+p} \hat{H}_g \ket{0}/2
  \\
   &  \ \ =
   u_{Q-p}u_{p}v_{Q-k}v_{k} \Big[ h(Q, p-k) + h(Q, Q-p-k) \Big] 
         +
        u_{Q-k}u_{k}v_{Q-p}v_{p} \Big[ h(Q, p-k) + h(Q, Q-p-k) \Big] 
        \\ &   \ \ +
         u_{Q-p}u_{Q-k}v_{p}v_{k} \Big[ h(p-k, Q) + h(p-k, Q-p-k) \Big] 
   +
        u_{k}u_{p}v_{Q-k}v_{Q-p} \Big[ h(p-k, Q) + h(p-k, Q-p-k) \Big]  
        \\ &   \ \ +
        u_{Q-p}u_{k}v_{Q-k}v_{p} \Big[ h(Q-p-k, Q) + h(Q-p-k, p-k) \Big] 
        +
        u_{Q-k}u_{p}v_{Q-p}v_{k} \Big[ h(Q-p-k, Q) + h(Q-p-k, p-k) \Big]
    \end{split}
\end{equation}

We now focus on the linear response induced by a density perturbation at a given momentum $Q$, $\hat{W} = \hat{n}(Q) - n \delta_{Q,0}  = \frac{1}{L}\sum_k \hat{b}^\dagger_{k+Q} \hat{b}_k
- n \delta_{Q,0}$ and probed at momentum $Q'$.
By defining $\hat{W}_g \equiv \hat{\mathcal{U}}_g^\dagger \hat{W} \hat{\mathcal{U}}_g$, the source terms are given by 
\begin{equation}
    f^-_{\beta}(Q')
     =
     \bra{0} \hat{b}_{Q'} 
     \hat{W}_g
     \ket{0} \\
 = \beta (u_Q + v_Q) \delta_{Q, Q'},
\end{equation}
\begin{equation}
    f^+_{\beta}(Q')
     =
     \bra{0} \hat{b}_{Q'} 
     \hat{W}_g^\dagger
     \ket{0} \\
 = \beta (u_Q + v_Q) \delta_{Q, -Q'},
\end{equation}

\begin{equation}
\begin{split}
    f^-_{\lambda_k}(Q') &=
     \frac{\bra{0} \hat{b}_{k} \hat{b}_{Q'-k}  \hat{W}_g \ket{0}
   }{
   \sqrt{1+\delta_{k,Q'-k}}}\\
   & = \frac{u_{Q-k}v_k + u_kv_{Q-k}}{\sqrt{1+\delta_{k,Q-k}}} \delta_{Q, Q'},
   \end{split}
\end{equation}
\begin{equation}
\begin{split}
    f^+_{\lambda_k}(Q') &=
     \frac{\bra{0} \hat{b}_{k} \hat{b}_{Q'-k}  \hat{W}_g^\dagger \ket{0}
   }{
   \sqrt{1+\delta_{k,Q'-k}}}\\
   & = \frac{u_{Q-k}v_k + u_kv_{Q-k}}{\sqrt{1+\delta_{k,Q-k}}} \delta_{Q, -Q'},
   \end{split}
\end{equation}
Notice that $f^{\mp}(Q) \propto \delta_{Q,\pm Q'}$ by momentum conservation. The susceptibility $\chi(Q, \omega)$ is defined as the coefficient of the $\e^{-i(\omega+i\eta) t}$ Fourier component of the linear response $\langle \hat{W}^\dagger (Q') \rangle = \vec{F}^\dagger \cdot \vec{X}(Q, \omega)$, leading to Eq. \eqref{eq:GaussDSF} in the main.

\twocolumngrid

\bibliography{bibliography}

\end{document}